\documentclass[notitlepage, 12pt]{article}

\usepackage{amssymb,amsmath,amsfonts,eurosym,geometry,ulem,graphicx,caption,color,setspace,sectsty,comment,footmisc,caption,natbib,pdflscape,array,hyperref,subcaption,threeparttable,threeparttablex, amsthm, rotating,chngcntr,pdflscape,subfloat, pifont, mathtools, csquotes, longtable}
\newcolumntype{L}{>{\centering\arraybackslash}m{1cm}}
\newcolumntype{C}[1]{>{\centering\let\newline\\arraybackslash\hspace{0pt}}m{#1}}
\newcolumntype{R}[1]{>{\raggedleft\let\newline\\arraybackslash\hspace{0pt}}m{#1}}

\begin{document}

\begin{titlepage}
\title{Social Media Emotions and Market Behavior\thanks{I am extremely grateful to Stefania Albanesi, Lee Dokyun, Sera Linardi, and Zhi Da for their continued guidance and support throughout this project. Special thanks to StockTwits for sharing their data. This research was supported in part by the University of Pittsburgh Center for Research Computing through the resources provided.}}

\author{Domonkos F. Vamossy\thanks{Department of Economics, University of Pittsburgh, d.vamossy@pitt.edu.}}
\date{\today}
\maketitle

\noindent

\begin{abstract}
	\singlespacing
I explore the relationship between investor emotions expressed on social media and asset prices. The field has seen a proliferation of models aimed at extracting firm-level sentiment from social media data, though the behavior of these models often remains uncertain. Against this backdrop, my study employs EmTract, an open-source emotion model, to test whether the emotional responses identified on social media platforms align with expectations derived from controlled laboratory settings. This step is crucial in validating the reliability of digital platforms in reflecting genuine investor sentiment. My findings reveal that firm-specific investor emotions behave similarly to lab experiments and can forecast daily asset price movements. These impacts are larger when liquidity is lower or short interest is higher. My findings on the persistent influence of sadness on subsequent returns, along with the insignificance of the one-dimensional valence metric, underscores the importance of dissecting emotional states. This approach allows for a deeper and more accurate understanding of the intricate ways in which investor sentiments drive market movements.

\vspace{0in}
 \bigskip
\end{abstract}

\setcounter{page}{0}
\thispagestyle{empty}
\end{titlepage}
\pagebreak \newpage

\doublespacing


\clearpage

\section{Introduction} \label{sec:introduction}

The intricate relationship between human emotions and markets is a foundational aspect of economic theory, highlighted by John Maynard Keynes's concept of ``animal spirits". He argued that it is the spontaneous optimism, not grounded in mathematical expectations, that often drives decisions:

\begin{quote}
    Even apart from the instability due to speculation, there is the instability due to the characteristic of human nature that a large proportion of our positive activities depend on spontaneous optimism rather than on a mathematical expectation, whether moral or hedonistic or economic. Most, probably, of our decisions to do something positive, the full consequences of which will be drawn out over many days to come, can only be taken as a result of animal spirits – of a spontaneous urge to action rather than inaction, and not as the outcome of a weighted average of quantitative benefits multiplied by quantitative probabilities. \cite{maynard}
\end{quote}

This quote captures the essence of the human element in economic dynamics, where sentiment and instinct can outweigh rational calculations. \cite{akerlof2010animal} further expanded on this in, arguing that understanding the emotional and psychological drivers is crucial for navigating financial markets effectively. They suggest that the economy's functioning is significantly influenced by confidence, fairness, corruption, and money illusion, all facets of Keynes's original ``animal spirits." 

In this paper, I examine the link between firm-specific investor emotions and the movements in their asset prices. My research is guided by two main questions: (1) Can firm-specific investor emotions predict daily price changes? and (2) Are the emotions derived from social media consistent with those observed in laboratory experiments? The challenge of accurately measuring investor emotions has led researchers to rely on indirect proxies or limit their analysis to experimental settings. However, by leveraging a large and unique dataset alongside modern advancements in text analysis, I have managed to navigate the complexities associated with studying investor emotions. Specifically, I utilize data from StockTwits, a social media platform where investors exchange views on stocks, which provides firm-specific posts, enabling the calculation of emotions. My analysis incorporates an extensive dataset comprising over 88 million messages collected from 2010 to September 2021.

This paper leverages \cite{vamossy2023emtract}, which employs deep learning techniques and an extensive dataset of investor communications. Specifically, the \cite{vamossy2023emtract} model processes text from social media and assigns to each message a set of variables that represent seven distinct emotional states: neutral, happy, sad, anger, disgust, surprise, and fear. This approach follows the emotional framework outlined by \cite{breaban2018emotional}. The emotional variables I work with are probabilistic, meaning that the sum of the seven emotions for any given message equals 1. For example, the text ``not feeling it :)" would yield an emotion tuple like (0.064, 0.305, 0.431, 0.048, 0.03, 0.038, 0.084), corresponding to the aforementioned emotional states. Changing the emoticon from :) to :( in ``not feeling it :(" alters the model's output significantly, demonstrating how the model adeptly integrates emoticons to refine the emotional analysis of the text.\footnote{For additional examples, visit the \href{https://huggingface.co/vamossyd/emtract-distilbert-base-uncased-emotion}{interactive interface}.} Using this tool, I can investigate whether investor emotions specific to a firm, captured before the market opens, can predict the firm's daily price movements. My approach involves applying the tool to quantify the emotional content of each message through textual analysis, and then averaging these results for each firm by day, distinguishing between messages at market open and close. Beyond measuring emotional content, I also categorize messages using two classification schemes: one that differentiates messages related to earnings, firm fundamentals, or stock trading from general discussions, and another that separates original content from reposted information.

To analyze whether these emotions can predict daily price movements, I employ a fixed effects model that leverages within-firm emotional variations. Using within-firm variation shuts down some mechanisms, such as the possibility that consistently high-return firms might elicit consistently positive investor emotions. By including firm-specific fixed effects, I can ensure that my findings aren't merely reflections of this. Additionally, I control for calendar day fixed effects to account for any broad market influences that might simultaneously affect all firms.

To address potential estimation concerns, I take several steps. For example, I focus on emotions expressed before the market opens — between 4:00 PM the previous day and 9:29 AM on the trading day — to establish a clear temporal separation between the emotions and the subsequent trading behavior. This helps in excluding reactionary emotions. To tackle the risk of misattribution, where the emotions captured might not accurately reflect the true sentiment, \cite{vamossy2023emtract} validate the emotion measures by comparing them with self-tagged bullish or bearish texts. I add to this by examining the link between emotion measures and contemporaneous asset prices and conducting robustness checks with an alternative emotion model. My findings suggest that assets that have increased in value tend to be associated with messages classified as happier, and these observations remain consistent even when using the alternative emotion model.

My analysis focuses on understanding how investor emotions influence daily asset price movements, leading to three key findings. First, I document that firm-specific investor emotions can predict the firm's daily price movements. For example, I find that variations in investor enthusiasm are associated with slightly higher daily returns. This effect is evident in both messages that provide new information and those that circulate existing information. When I compare messages related to earnings, firm fundamentals, or stock trading with those containing other types of information, the latter seems to have a slightly larger impact on daily returns. To confirm the relationship between daily price movements and investor emotions, I also explore alternative emotion variables from emotion metadata compiled by other researchers. My findings remain significant, showing comparable point estimates. Moreover, I observe that the effects of emotions are more pronounced during periods of elevated short interest and lower liquidity. Second, emotions expressed on social media align with those expected from controlled laboratory experiments. This consistency strengthens the case for using social media as a viable proxy for investor sentiment, indicating that the emotional expressions captured online are indicative of wider investor behavior patterns. Last, the persistence of sadness on subsequent returns and the insignificance of the combined valence metric underscores the importance of dissecting emotional states to gain a deeper and more accurate insight into how investor sentiments drive market movements.

My research contributes to the behavioral finance field by extending the understanding of the link between market behavior and emotional states. Unlike previous studies that often relied on indirect proxies, such as \cite{kamstra2003winter} linking seasonal mood changes to market returns or \cite{hirshleifer2003good} correlating weather with stock performance, my approach directly measures investor emotions through social media posts related to specific stocks, addressing concerns about the relevance of emotional measures to the firms. This direct method of capturing investor sentiment allows for a clearer connection between emotions and stock prices, sidestepping potential confounders highlighted by critiques like \cite{jacobsen2008weather}. While research in this area has predominantly focused on market-level emotions and index returns\footnote{\cite{bollen2011twitter} discovered that the overall mood on Twitter could forecast movements in the stock market, whereas \cite{gilbert2010widespread} identified that the degree of anxiety in Live Journal blog posts was indicative of forthcoming declines in stock prices.}, with few exceptions like \cite{li2016can} and \cite{long2021just} exploring firm-specific sentiments, my work adds a novel dimension by showing that investor emotions at the firm level can predict daily stock price movements, thereby offering a more granular perspective on how sentiment influences market dynamics.

The connection between market behavior and emotional states has been extensively examined in controlled laboratory settings, where researchers have investigated how emotions contribute to the formation of bubbles in experimental asset markets. For example, \cite{breaban2018emotional} utilized facial expressions of traders to measure emotions, finding that positive emotions correlate with higher prices and bigger bubbles, whereas pre-market fear among traders tends to result in lower prices. \cite{andrade2016bubbling}  also observed that bubbles tend to be larger when induced investor enthusiasm is higher. My contribution to this field is demonstrating that emotions reflected in investor messages on social media platforms mirror those observed in laboratory experiments, suggesting a parallel in how emotions influence market dynamics both in controlled settings and in the real world via observational data. Nevertheless, it's crucial to recognize that the observational nature of social media data, while insightful, lacks the precision and control over variables that laboratory experiments offer. This inherent limitation complicates the task of drawing clear causal inferences from the data observed in such environments.

Another contribution of my paper is to the body of research examining social media's impact on capital markets. The relevance of social media in this domain is underscored by studies such as \cite{blankespoor2014role}, who demonstrate that firms can diminish information asymmetry among investors by using Twitter to widely distribute news, press releases, and other disclosures. \cite{jung2018firms} report that a significant portion of S\&P 1500 firms have established a corporate presence on social media platforms like Facebook and Twitter. Additionally, there's a growing body of work, including research on StockTwits and Twitter, that investigates how investors' engagement with various online platforms, from search engines and financial websites to forums, influences market dynamics. This line of inquiry has yielded mixed results on the predictive power of online information for future earnings and stock returns. For instance, \cite{da2011search} use Google search volume as an indicator of investors' information demand, finding that increased searches foreshadow short-term stock price increases followed by a reversal within a year. \cite{drake2012investor} observe that the correlation between returns and earnings weakens when Google searches spike before earnings announcements. Studies by \cite{antweiler2004all} and \cite{das2007yahoo} link the volume of message board posts to stock return volatility but not to the returns themselves. \cite{chen2014wisdom} show that user-generated content on the Seeking Alpha platform can forecast earnings and long-term stock returns post-report publication. The literature also delves into how social media activity around earnings announcements affects investor behavior, with \cite{curtis2014investor} finding a connection between social media buzz and heightened sensitivity to earnings news and surprises, and  \cite{cookson2020don} indicating that while StockTwits discourse may not directly impact market movements, the platform's message disagreements are a reliable predictor of unusual trading volumes.\footnote{Additional studies by \cite{vamossy2023social} and \cite{vamossyemotions}.}

The remainder of the paper is organized as follows. Section \ref{sec:data} describes the data; Section \ref{sec:primary} presents my primary results; Section \ref{sec:additional} explores heterogeneous effects and conducts a sensitivity analysis. Section \ref{sec:conclusion} concludes.

\section{Data}\label{sec:data}

\subsection{Social Media Data (StockTwits)}

My dataset on investor emotions is sourced from StockTwits, established in 2008 as a platform for investors to exchange opinions on stocks. The interface of StockTwits is reminiscent of Twitter, allowing users to post messages limited to 140 characters, which was expanded to 280 characters in late 2019. Users employ ``cashtags" followed by a stock's ticker symbol (for example, \$AMZN) to associate their messages with specific companies. While StockTwits does not offer direct integration with other social media networks, it does enable users to share their posts on personal accounts across Twitter, LinkedIn, and Facebook.

My initial dataset spans from 1 January 2010 to 30 September 2021, encompassing a total of 242 million messages. For each message, I can access sentiment indicators marked by users (labeled as bullish, bearish, or unclassified), a sentiment score calculated by StockTwits, ``cashtags" linking messages to specific stocks, the number of likes a message receives, and a unique user identifier. This identifier enables me to delve into user characteristics, such as the number of followers they have. For the majority of users, there is also data on their self-declared investment philosophy, which can be categorized in two ways: (1) by their Approach - including technical, fundamental, momentum, value, growth, and global macro strategies; and (2) by their Holding Period - classified as a day trader, swing trader, position trader, or long-term investor. To analyze variations among investors, I combine technical with momentum and value with fundamental approaches and differentiate between short- and long-term investment horizons. Additionally, users disclose their level of experience, which ranges from novice, through intermediate, to professional. This detailed information on users' investment styles, experience levels, and strategies is instrumental in examining the diversity of investor emotions on the platform.

I eliminate messages that seem to be automated, defining these as any cleaned message posted more than 100 times by the same user throughout the period from 2010 to 2021. My focus is on messages explicitly linked to specific stocks, leading me to concentrate on those mentioning only one ticker symbol. Additionally, I set a threshold requiring a minimum of ten posts per stock for each trading session (including both market and non-market hours) to minimize the impact of random noise. I also exclude inactive tickers as defined by SECSTAT and limit my study to common ordinary shares listed on U.S. exchanges. These sample restrictions are detailed in Table \ref{tab:sample_restrictions}. I end up with 88M messages authored by 984,434 users, covering 4,319 tickers.

\begin{figure}[!h]
	\centering
	\begin{subfigure}[b]{0.49\textwidth}
		\includegraphics[scale=0.4]{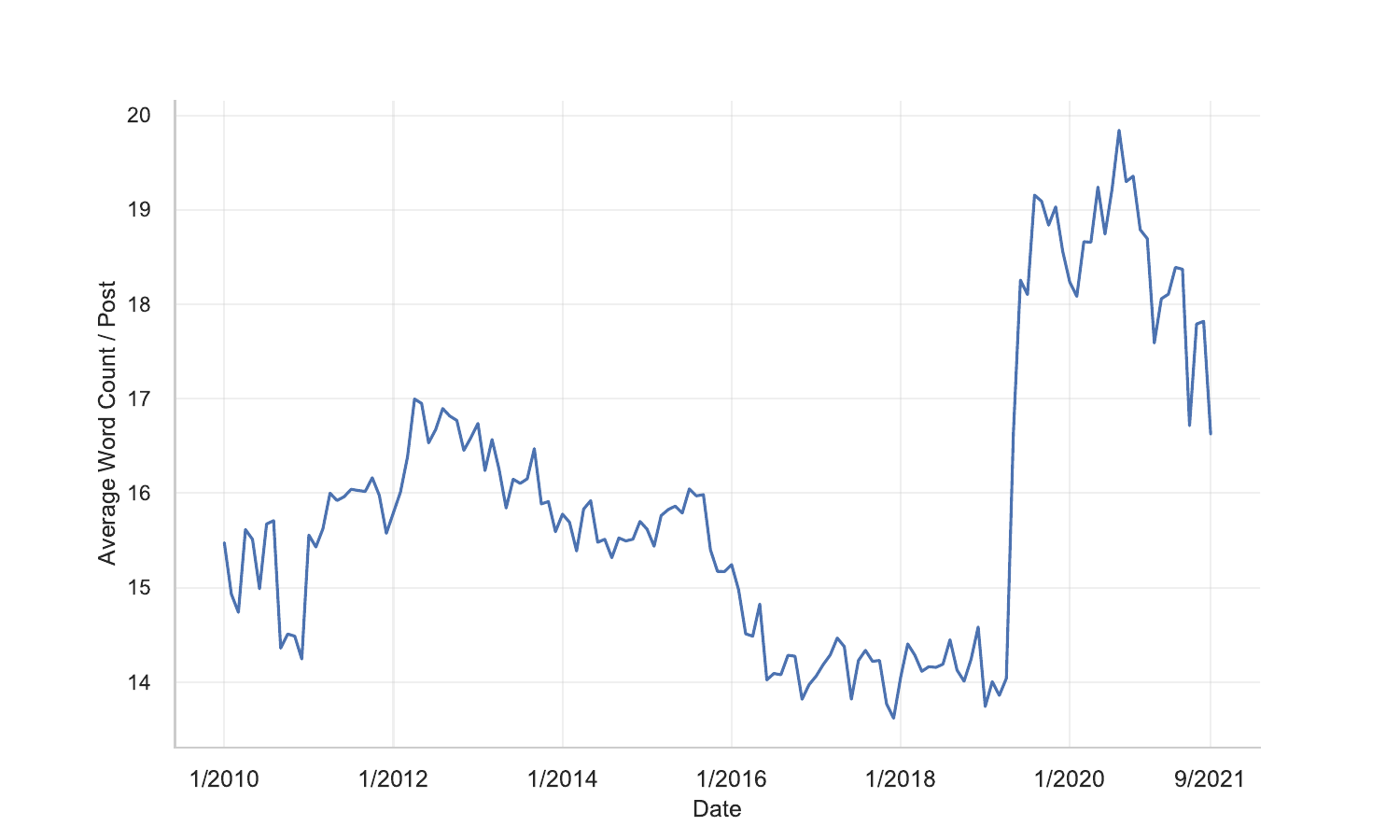}
		\caption{}
	\end{subfigure}
	\begin{subfigure}[b]{0.49\textwidth}
		\includegraphics[scale=0.4]{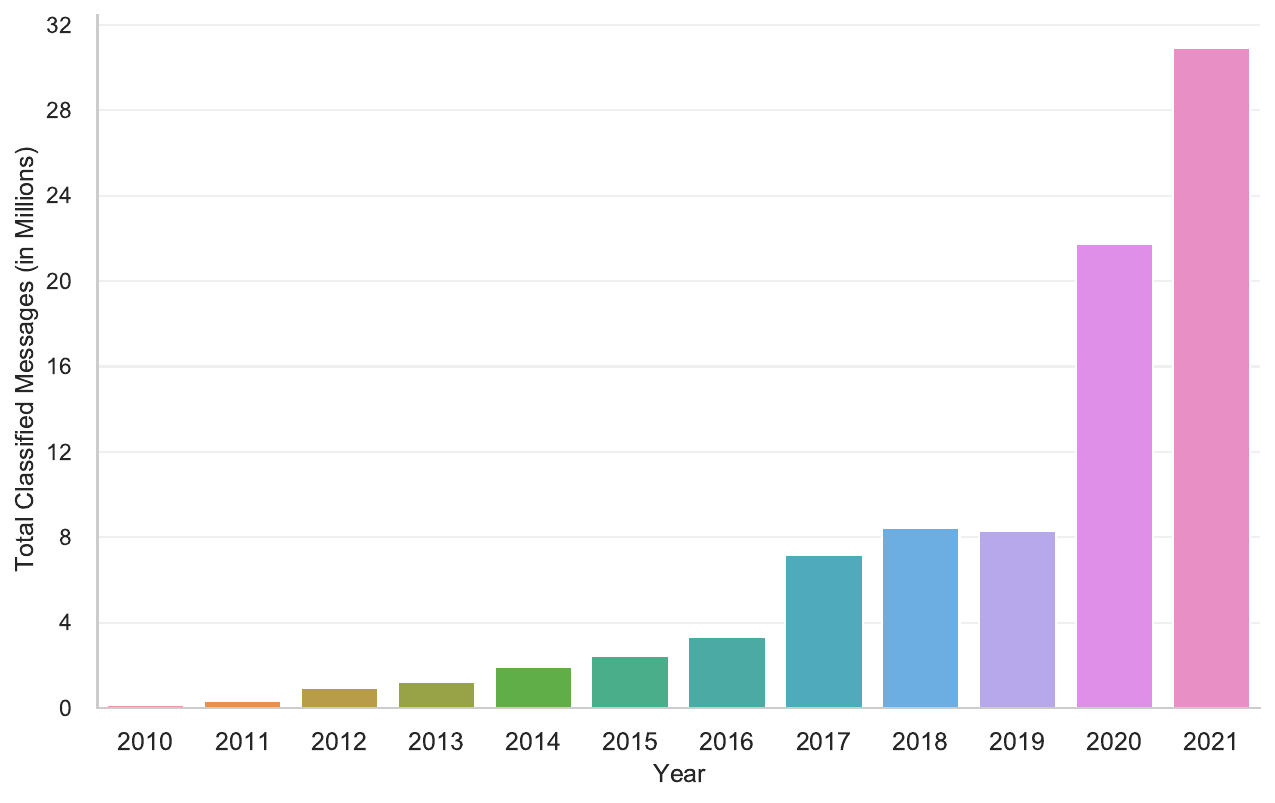}
		\caption{}
	\end{subfigure}
		
	\begin{subfigure}[b]{0.49\textwidth}
		\includegraphics[scale=0.4]{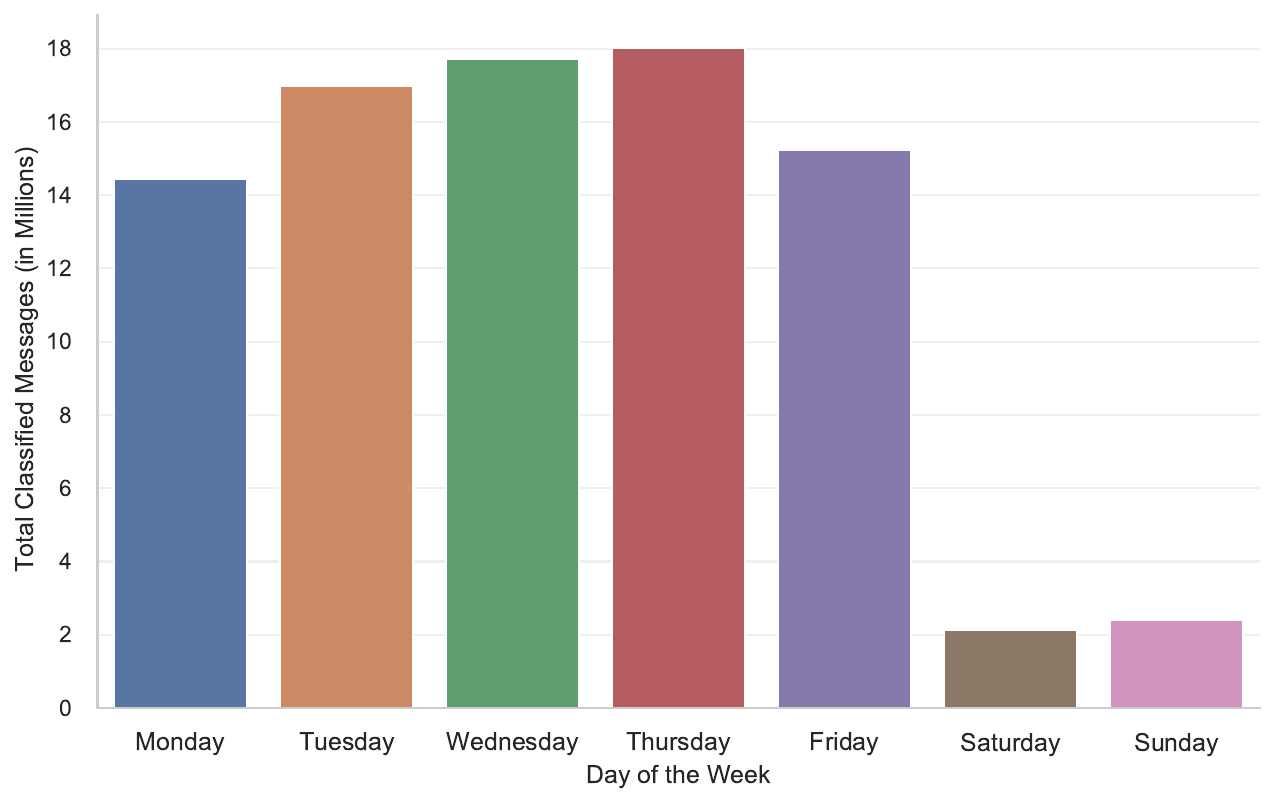}
		\caption{}
	\end{subfigure}
	\begin{subfigure}[b]{0.49\textwidth}
		\includegraphics[scale=0.4]{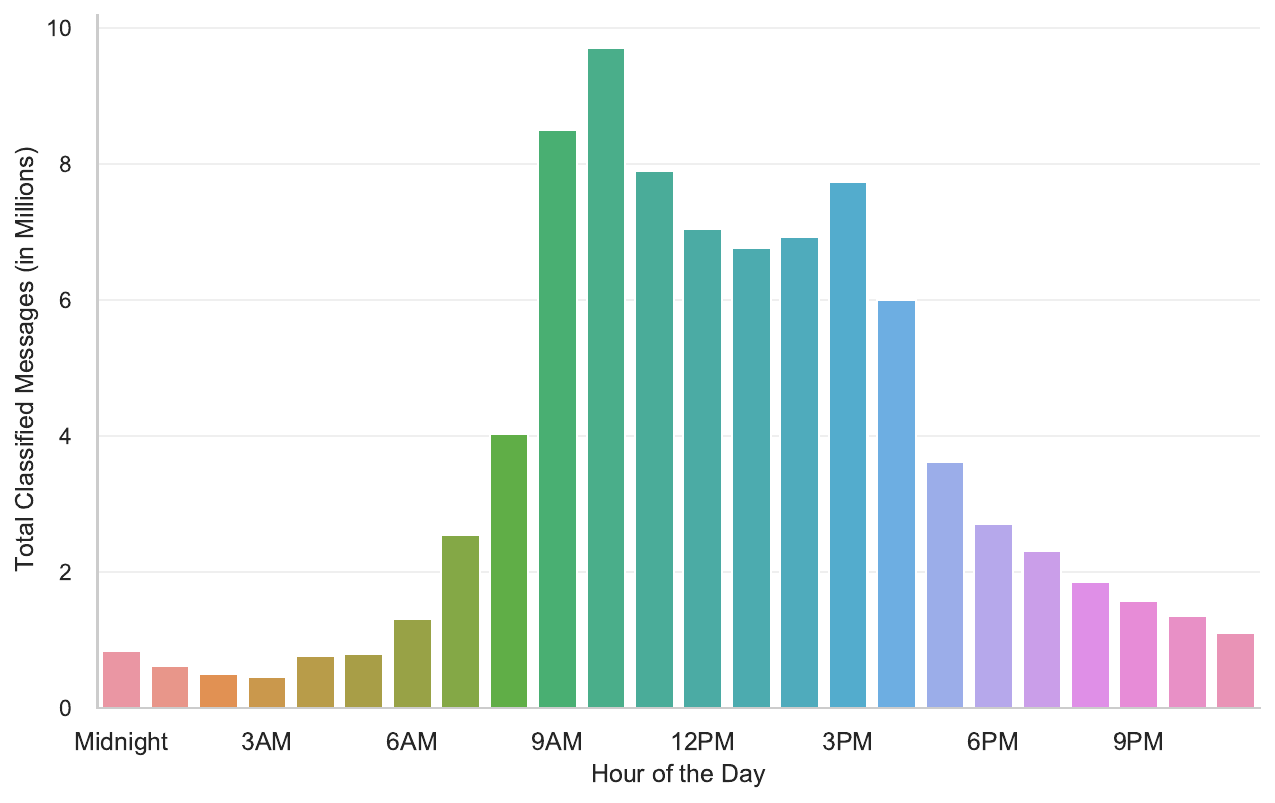}
		\caption{}
    \end{subfigure}	

	\begin{subfigure}[b]{0.49\textwidth}
		\includegraphics[scale=0.4]{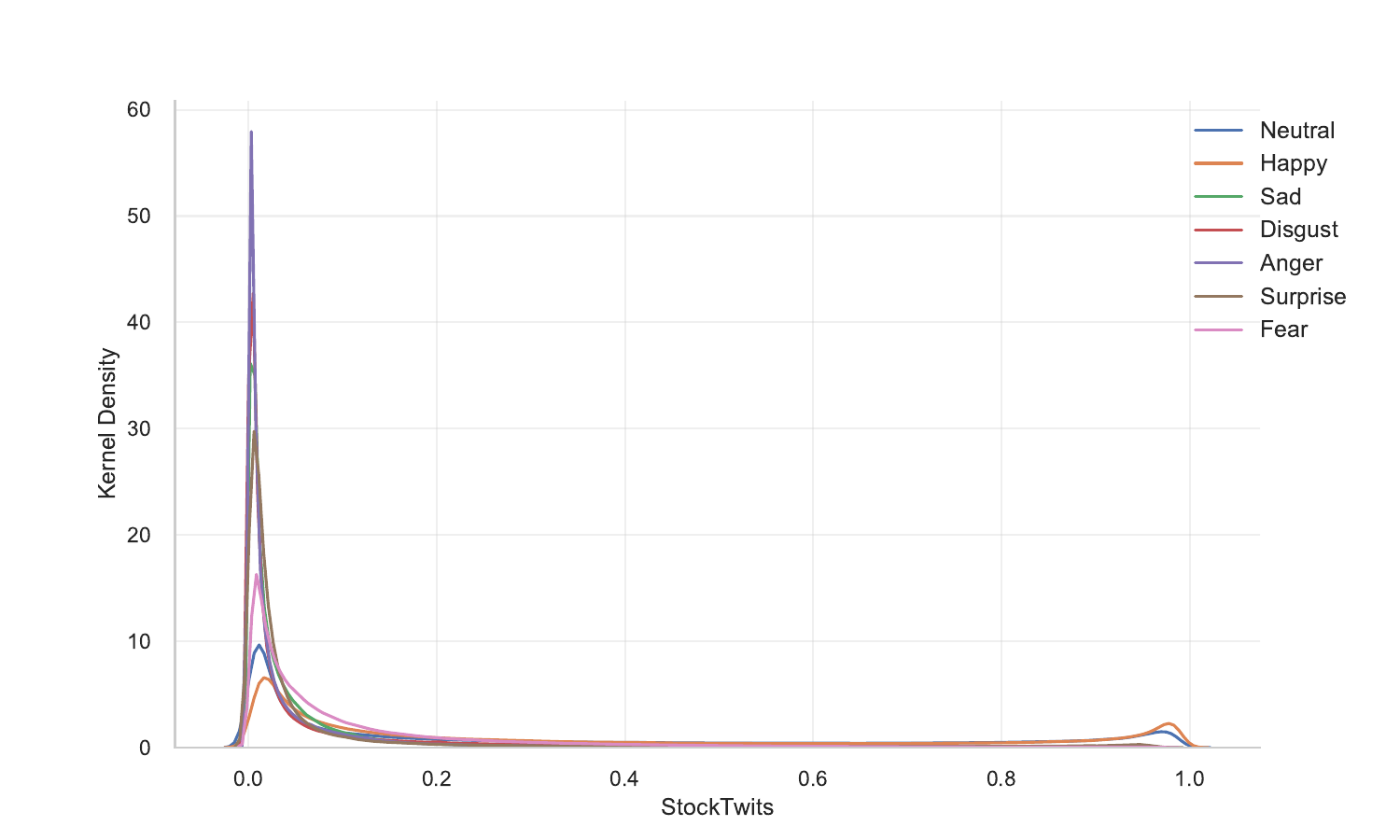}
		\caption{}
    \end{subfigure}	

	\caption{Post Characteristics}\label{fig:post_distributions}
		\begin{flushleft}
		 \footnotesize{Notes: This includes my final dataset for pre-market and market sessions. In Panel (a), I chart the average word count per message over time, and Panel (b) illustrates the trend in the volume of posts in my dataset. Panel (c) presents the distribution of posts by day of the week, Panel (d) shows the distribution by hour of the day, and Panel (e) plots the distribution of classified messages.}
		\end{flushleft}
\end{figure}

In Panel (a) of Figure \ref{fig:post_distributions}, I plot the average word count of messages over time, noting a consistent trend with a notable increase in late 2019 due to the expansion of the character limit from 140 to 280. The average message length reaches a high of 20 words, and 97.5\% of the messages in my dataset contain 64 words or fewer. Given that I use only the first 64 words of each message to identify the underlying emotions, this cap is unlikely to compromise the accuracy of my analysis.

In Panel (b) of Figure \ref{fig:post_distributions}, I present the trend of message volume over time in my dataset, showing a significant increase in the early stages, a leveling off around 2017, and then a marked rise in 2020. To account for the evolving nature of both the sample size and the content of posts, I incorporate time fixed effects in my analysis.

I further investigate the timing of investors' posts, specifically whether they align with the influx of daily news, reflecting real-time adjustments in beliefs, or if they tend to occur in the evening, suggesting a more reflective, analytical approach after work hours. In Panels (c) and (d) of Figure \ref{fig:post_distributions}, I display the distribution of messages by both the day of the week and the hour of the day. The bulk of activity on the platform coincides with market hours (Monday to Friday, 9 AM to 4 PM), indicating that investors are likely reacting and adjusting their perspectives in response to live market developments.

\subsection{Short Interest and Pricing}
I constructed additional variables using conventional data sources. Variables related to stock prices were sourced from the merged CRSP/COMPUSTAT file, while short interest data came from Compustat's Supplemental Short Interest File. I then matched this information with the StockTwits data based on the ticker symbols and dates.

\subsection{Descriptive Statistics}
Table \ref{tab:summary_stats} presents the descriptive statistics for the variables used in my analysis, including the mean, standard deviation (shown in parentheses), and the within-firm standard deviation (indicated by brackets). In Panel A, where I summarize the statistics for variables derived from social media, a significant portion of the messages, averaging 51.1\%, are classified as neutral, with happy messages constituting 24.6\%. Analyzing the distribution of emotion variables reveals a tendency towards positive skewness, indicating that investors might be more inclined to express optimism rather than pessimism on social media platforms. Fear and surprise emerge as the third and fourth most prevalent emotions, respectively, followed by disgust, sadness, and anger. Similar to the return variables, the within-firm standard deviation for the emotion variables is only marginally lower than the overall sample standard deviation, suggesting substantial variability that can be utilized even when applying a firm-date fixed effect model. I plot the distribution of emotions in Panel (e) of Figure \ref{fig:post_distributions}.

To provide context for the scale of my other variables, I compare my final estimation dataset with a parallel sample from CRSP/Compustat for the same period, adhering to identical sample criteria (i.e., active, common ordinary shares on U.S. exchanges). The findings, presented in Panel B of Table \ref{tab:summary_stats}, reveal several notable trends. First, it appears that StockTwits users are more inclined to talk about stocks that have recently increased in value or are currently on the rise: the mean past monthly return in my dataset is 3.3 percentage points higher, the close-to-open return is up by 0.3 percentage point, and the one-day lagged open-to-close return is 0.1 percentage point above those in the CRSP dataset. Secondly, these discussed stocks exhibit a 0.3 percentage point lower open-to-close return on average, hinting at potential mean-reversion. Additionally, I observe a heightened interest among social media investors in firms characterized by larger trading volumes, higher volatility, greater market capitalization, and increased short interest, but with lesser institutional ownership and lower historical returns.

Table \ref{tab:corr_matrix} showcases the pairwise correlation coefficients for the variables in my analysis. These include my emotion variables, daily returns, and key control variables such as the lagged open-close return, close-open return, trading return over the past 20 days, and volatility over the past 183 days. The emotion measures show statistically significant correlations with daily returns and among themselves. Specifically, the neutral emotion category exhibits the highest correlations with both positively and negatively valenced emotions, serving as the baseline group, and hence omitted in the regression specifications. Additionally, to address the issue of a high correlation between certain negative emotions, such as anger and disgust, these can be merged into a single ``hate" category, though this adjustment does not markedly influence the findings. The primary focus remains on the emotions of happiness and fear, which are crucial for understanding the market dynamics in terms of the classic dichotomy of fear and greed. I considered incorporating additional control variables, but variables like trading volume and market capitalization were highly correlated with volatility, leading to their exclusion to prevent multicollinearity. Although short interest and institutional ownership were not included in the main models due to their potential to significantly reduce the sample size, I examine these factors separately in Table \ref{tab:pre_market_volat}.

\begin{table}[htbp]\centering
\def\sym#1{\ifmmode^{#1}\else\(^{#1}\)\fi}
\begin{threeparttable}
\small
\caption{Itemized Sample Restrictions}\label{tab:sample_restrictions}
\begin{tabular}{lll}
 &  & Messages \\ \hline 
   &  &  \\
StockTwits Data 2010-2021* &  & 242,362,477
 \\ \hline
     &  &  \\
 Keep & &  \\ \hline 
   &  &  \\
Single Ticker
 &  & 180,298,172
 \\
Not Automated
&  & 167,040,745
 \\ 
Active, Common Ordinary Shares, Traded on US Exchanges**
&  & 103,248,233
 \\ 
At Least 10 Messages during Non-Market/Market Hours
&  & 88,055,174

 \\ \hline 
   &  &  \\
Final Pre-Market (4pm-9am) Sample & & 37,657,214          \\ 
   & & \\
Final Market (9am-4pm) Sample & & 50,397,960 \\ 
   & & \\ \hline \hline 
\end{tabular}
\begin{tablenotes}
\footnotesize 
\item Notes: \sym{*}Sample starts January 1st 2010 and runs till September 30th 2021. \sym{**}I filter out stocks for which the security status (SECSTAT is ``I") is inactive and restrict my sample to common ordinary shares (TPCI is ``0") traded on US exchanges (EXCHG is 11, 12, 14, or 17).
\end{tablenotes}
\end{threeparttable}
\end{table}

\begin{table}[htbp]\centering
\begin{threeparttable}
\small 
\def\sym#1{\ifmmode^{#1}\else\(^{#1}\)\fi}
\caption{Summary Statistics\label{tab:summary_stats}}
\begin{tabular}{l*{3}{c}}
\hline\hline
& & &\\[\dimexpr-\normalbaselineskip+2pt]
                    &\multicolumn{1}{c}{CRSP/Compustat-StockTwits}&\multicolumn{1}{c}{CRSP/Compustat}&\multicolumn{1}{c}{Difference} \\
\hline
& & &\\[\dimexpr-\normalbaselineskip+2pt]
\multicolumn{4}{l}{\textbf{Panel A: Social Media Information}} \\
& & &\\[\dimexpr-\normalbaselineskip+2pt]
\textcolor{white}{...}Happy               &       0.246& &\\
                    &     (0.150) [0.108] & &\\
& & &\\[\dimexpr-\normalbaselineskip+2pt]
\textcolor{white}{...}Sad                 &         0.036&     &   \\
                    & (0.0294) [0.0257]  &   &\\
& & &\\[\dimexpr-\normalbaselineskip+2pt]
\textcolor{white}{...}Fear                &        0.085&   &    \\
                    & (0.0532) [0.0432]   &   &\\
& & &\\[\dimexpr-\normalbaselineskip+2pt]
\textcolor{white}{...}Disgust             &        0.037&    &   \\
                    &  (0.0343) [0.0284]  &   &\\
& & &\\[\dimexpr-\normalbaselineskip+2pt]
\textcolor{white}{...}Anger               &        0.024&    &   \\
                    &  (0.0213) [0.0174] &   &\\
& & &\\[\dimexpr-\normalbaselineskip+2pt]
\textcolor{white}{...}Surprise            &        0.062&   &    \\
                    &  (0.0511) [0.0438]  &   &\\
& & &\\[\dimexpr-\normalbaselineskip+2pt]
\textcolor{white}{...}Neutral             &        0.511&   &    \\
                    &   (0.248) [0.155]  &   &\\
& & &\\[\dimexpr-\normalbaselineskip+2pt]
\multicolumn{4}{l}{\textbf{Panel B: Financial Information}} \\
& & &\\[\dimexpr-\normalbaselineskip+2pt]
\textcolor{white}{...}Open-Close Return  & -0.002    & 0.000   & -0.003\sym{***}  \\
                    &  (0.0508) [0.0481] & (0.032)  &\\
& & &\\[\dimexpr-\normalbaselineskip+2pt]
\textcolor{white}{...}Close-Open Return &     0.004   &  0.001    &      0.003\sym{***}\\
            &   (0.0410) [0.0390]  &  (0.022)  &                     \\
& & &\\[\dimexpr-\normalbaselineskip+2pt]
\textcolor{white}{...}Open-Close Return$_{-1}$      &   0.001    &     0.000  &      0.001\sym{***}\\
            &   (0.0536) [0.0507] &   (0.032)  &                     \\
& & &\\[\dimexpr-\normalbaselineskip+2pt]
\textcolor{white}{...}Return$_{-20,-1}$        &       0.047 &      0.014&           0.033\sym{***}   \\
                    &  (0.297) [0.267] & (0.158)   &\\
& & &\\[\dimexpr-\normalbaselineskip+2pt]
\textcolor{white}{...}\$ Volume$_{-183,-1}$ &     17.032&   15.229 &   1.803\sym{***}   \\
                    &   (2.615) [0.986]  & (2.764)  &\\
& & &\\[\dimexpr-\normalbaselineskip+2pt]
\textcolor{white}{...}Volatility$_{-183,-1}$    &    0.039&   0.027&      0.012\sym{***}      \\
                    & (0.0217 [0.00932]  &  (0.017)  &\\
& & &\\[\dimexpr-\normalbaselineskip+2pt]
\textcolor{white}{...}Market Cap$_{-1}$     &       21.398&    20.742&      0.655\sym{***}      \\
                    &  (2.624) [0.921]  & (2.076) &\\
& & &\\[\dimexpr-\normalbaselineskip+2pt]
\textcolor{white}{...}Institutional Ownership&          0.518&   0.602&     -0.084\sym{***}   \\
                    &   (0.322) [0.106] & (0.318) &\\
& & &\\[\dimexpr-\normalbaselineskip+2pt]
\textcolor{white}{...}Short Interest    &         0.074&    0.042&   0.032\sym{***}    \\
                    & (0.0819) [0.0478]  &   (0.053) &\\
\hline
& & &\\[\dimexpr-\normalbaselineskip+2pt]
 Observations       & 479,463 &       10,428,859  &            \\
\hline\hline
\end{tabular}
   \begin{tablenotes}
     \footnotesize
    \item Notes: \$ Volume and Market capitalization are in logs. Institutional Ownership, Short Interest, Return, and Volatility are in units. Close-Open Return refers to the return between closing price at $t-1$ to the opening price at $t$. \sym{*} \(p<0.10\), \sym{**} \(p<0.05\), \sym{***} \(p<0.01\). Standard deviations in parentheses, within-firm standard deviation in brackets for my CRSP/Compustat-StockTwits sample only (I remove firm and date fixed effects to help interpret effect sizes). 
   \end{tablenotes}
\end{threeparttable}
\end{table}
 
\begin{sidewaystable}[htbp]\centering
\begin{threeparttable}
\footnotesize
\caption{Correlation Matrix}\label{tab:corr_matrix}
\def\sym#1{\ifmmode^{#1}\else\(^{#1}\)\fi}
\begin{tabular}{l*{12}{c}}
\hline\hline
                &\multicolumn{12}{c}{}                                                                                                                \\
                &Daily Return         &    Happy         &      Sad         &     Fear         &  Disgust         &    Anger         & Surprise         &  Neutral         &   Return$_{oc, -1}$         &Return$_{co}$           &Return$_{-20,-1}$                 \\
   &    &              &                  &                  &                  &                  &                  &                  &                  &                  &                  &                  \\
\hline
   &    &              &                  &                  &                  &                  &                  &                  &                  &                  &                  &                  \\
                &                  &                  &                  &                  &                  &                  &                  &                  &                  &                  &                  &                  \\
Happy           &    -0.03\sym{***}&              &                  &                  &                  &                  &                  &                  &                  &                  &                  &                  \\
                &                  &                  &                  &                  &                  &                  &                  &                  &                  &                  &                  &                  \\
Sad             &    -0.02\sym{***}&     0.26\sym{***}&              &                  &                  &                  &                  &                  &                  &                  &                  &                  \\
                &                  &                  &                  &                  &                  &                  &                  &                  &                  &                  &                  &                  \\
Fear            &    -0.02\sym{***}&     0.33\sym{***}&     0.59\sym{***}&              &                  &                  &                  &                  &                  &                  &                  &                  \\
                &                  &                  &                  &                  &                  &                  &                  &                  &                  &                  &                  &                  \\
Disgust         &    -0.02\sym{***}&     0.30\sym{***}&     0.52\sym{***}&     0.48\sym{***}&              &                  &                  &                  &                  &                  &                  &                  \\
                &                  &                  &                  &                  &                  &                  &                  &                  &                  &                  &                  &                  \\
Anger           &    -0.02\sym{***}&     0.34\sym{***}&     0.53\sym{***}&     0.51\sym{***}&     0.79\sym{***}&              &                  &                  &                  &                  &                  &                  \\
                &                  &                  &                  &                  &                  &                  &                  &                  &                  &                  &                  &                  \\
Surprise        &    -0.02\sym{***}&     0.32\sym{***}&     0.36\sym{***}&     0.38\sym{***}&     0.40\sym{***}&     0.42\sym{***}&              &                  &                  &                  &                  &                  \\
                &                  &                  &                  &                  &                  &                  &                  &                  &                  &                  &                  &                  \\
Neutral         &     0.03\sym{***}&    -0.85\sym{***}&    -0.60\sym{***}&    -0.67\sym{***}&    -0.64\sym{***}&    -0.66\sym{***}&    -0.62\sym{***}&              &                  &                  &                  &                  \\
                &                  &                  &                  &                  &                  &                  &                  &                  &                  &                  &                  &                  \\
Return$_{oc, -1}$           &    -0.03\sym{***}&     0.11\sym{***}&    -0.12\sym{***}&    -0.10\sym{***}&    -0.11\sym{***}&    -0.10\sym{***}&    -0.05\sym{***}&     0.00         &              &                  &                  &                  \\
                &                  &                  &                  &                  &                  &                  &                  &                  &                  &                  &                  &                  \\
Return$_{co}$     &    -0.07\sym{***}&     0.09\sym{***}&    -0.11\sym{***}&    -0.08\sym{***}&    -0.07\sym{***}&    -0.06\sym{***}&    -0.03\sym{***}&    -0.01\sym{*}  &     0.03\sym{***}&              &                  &                  \\
                &                  &                  &                  &                  &                  &                  &                  &                  &                  &                  &                  &                  \\
Return$_{-20,-1}$    &    -0.02\sym{***}&     0.12\sym{***}&    -0.06\sym{***}&    -0.02\sym{***}&    -0.04\sym{***}&    -0.03\sym{***}&    -0.00         &    -0.05\sym{***}&     0.20\sym{***}&    -0.00         &              &                  \\
                &                  &                  &                  &                  &                  &                  &                  &                  &                  &                  &                  &                  \\
Volatility$_{-183,-1}$&    -0.04\sym{***}&     0.42\sym{***}&     0.17\sym{***}&     0.16\sym{***}&     0.26\sym{***}&     0.28\sym{***}&     0.24\sym{***}&    -0.42\sym{***}&     0.02\sym{***}&     0.06\sym{***}&     0.16\sym{***}&              \\
                &                  &                  &                  &                  &                  &                  &                  &                  &                  &                  &                  &                  \\
\hline

Observations    &   479463         &                  &                  &                  &                  &                  &                  \\
\hline\hline
\end{tabular}
\begin{tablenotes}
\scriptsize 
\item Notes: Pairwise correlation with Bonferroni corrections. Continuous variables winsorized at the 0.1\% and 99.9\% levels. \textit{t} statistics in parentheses. \sym{*} \(p<0.05\), \sym{**} \(p<0.01\), \sym{***} \(p<0.001\). Return$_{co}$ refers to return between the closing price at $t-1$ and the opening price at $t$. Return$_{oc}$ refers to the open-close return.
\end{tablenotes}
\end{threeparttable}
\end{sidewaystable}

\clearpage

\section{Primary Findings}\label{sec:primary}

As demonstrated by \cite{vamossy2023emtract}, emotions can vary considerably over time. In this section, I delve into how such emotional shifts impact stock returns. I hypothesize that heightened investor enthusiasm could lead to increased buying activity, temporarily elevating stock prices. I explore this effect on daily price movements. 

Before delving into the main findings, I must clarify that my analysis does not establish a causal relationship between emotions on social media and stock price changes. I demonstrate that emotions extracted from from social media offer insights into stock valuation that are not covered by static stock characteristics, temporal patterns, or recent price trends. Moreover, this approach supports the insights gained from controlled lab experiments regarding the influence of investor emotions on trading behaviors.

\subsection{Empirical Framework}

My return regressions exploit within-firm variation. That is, I estimate the following model:
\begin{equation}\label{eq:emotion}
 Y_{it} = \alpha + \sum_{j=0}^{j=5} \beta_j \textrm{Emotion}_{jit} +\gamma Y_{i,t-1} + \zeta X_{i, t-1} + \delta_t + \delta_i + \epsilon_{it} 
\end{equation}

where $\textrm{Emotion}{jit}$ represents the emotion metric for firm i on date t, calculated from messages posted during non-market hours before the market opens. In the regression framework, neutral emotions serve as the reference category, with $j$ encompassing emotions such as happy, sad, anger, disgust, surprise, and fear. The primary dependent variable in most of my analysis is the daily open-close return. Day fixed effects ($\delta_t$) are included to account for significant daily news events or overall market returns, while firm fixed effects ($\delta_i$) help control for inherent characteristics of firms that might influence results, such as a firm's tendency to yield high returns possibly eliciting more excitement among investors before market opening. To address autocorrelation, I control for the previous day's open-close return. Additionally, I incorporate the close-open return (from the previous day's close to the current day's open) to verify that the emotions captured offer new information beyond after-hours and pre-market price changes. My models also factor in controls ($X_{i, t-1}$) for the return over the past 20 trading days and volatility. To dispel doubts that the emotion measure merely reflects shifts in trading volume, I base my dependent variable, the day t open-close return, on emotions from messages posted before the market opened on day t, ensuring clear temporal precedence of the emotion metric over the dependent variable.

In further refining my analysis, I categorize messages into two groups and calculate the emotion score for each. The first group consists of messages that directly reference a company's fundamentals or stock price, which I label as ``finance" messages. The second group, termed ``chat" messages, encompasses more general discussions. This distinction aids in shedding light on the influence of emotions within varied informational contexts. Additionally, I differentiate messages based on their content: ``original" messages are those that provide new information, characterized by not being a retweet and lacking hyperlinks, while ``dissemination" messages are those that circulate existing information. This classification helps in understanding the impact of emotions in messages offering new insights versus those that simply share what is already known.

\subsection{Testing Laboratory Experiments}

I first delve into whether the outcomes observed in controlled laboratory experiments about investor emotions and their trading behaviors are mirrored in the investor emotion data I have collected. Table \ref{tab:tests} categorizes the findings of pivotal laboratory experiments into directly and indirectly testable outcomes. To refine the reliability of these insights, I demean each metric by firm and date, aside from Finding III. By doing so, I close off certain mechanisms that could otherwise obscure the true relationship between emotions and market behavior.

In Panel A, I confirm the negative correlation between pre-market fear and subsequent price movements, in line with \cite{breaban2018emotional}. This is further substantiated by the observed correlations of fear, anger, and happiness with contemporaneous price movements, echoing the findings of both \cite{hargreaves2011emotions} and \cite{breaban2018emotional}. Next, I extend the comparison to indirectly testable hypotheses, as delineated in Panel B of Table \ref{tab:tests}. To test these, I operate under two key assumptions: first, that traders exhibiting loss aversion tend to steer clear of firms with smaller market capitalizations, possibly due to perceived higher risks; and second, that there is a relationship between trading volume and cash holdings, where increased trading activity suggests lower cash reserves among traders. My findings in Panel B replicate the findings of lab experiments.

Despite noting somewhat modest correlation coefficients, I validated the conclusions of two laboratory experiments that link emotions to trading actions. This comparison not only reinforces the relevance of emotions in trading decisions but also highlights the potential of social media data to reflect complex emotional dynamics observed in more controlled experimental settings.

\subsection{Emotions, Information Content, and Returns}

I next explore the connection between investor emotions and daily stock returns through my regression framework specified in \ref{eq:emotion}. Closest to my setting, lab evidence finds a positive (negative) relationship between investor enthusiasm (fear) before the market opens and subsequent returns (\cite{breaban2018emotional}). I empirically test this relationship in Table \ref{tab:pre_market}. In Column (1), I incorporate only my control variables, along with firm and date fixed effects. By directly comparing Column (1) to Column (2), I observe that emotions prior to market opening explain a small portion of the variance in daily returns, with an increase in $R^2$ by 0.05 percentage points. Additionally, I find that a one standard deviation increase in enthusiasm before the market opens on day t correlates with a 0.7\% increase in the standard deviation of daily stock returns. In Column (3), my results indicate that the influence of investor emotions diminishes for larger firms and those that are easier to value, such as those within the S\&P Super Composite index. My findings in Column (4) reveal more significant effects for stocks with higher user engagement (those with at least 100 messages), where a one standard deviation increase in pre-market happiness is associated with a 3.1\% increase in the standard deviation of daily open-close stock returns. These insights suggest that investor emotions, as derived from StockTwits, provide valuable information relevant to stock valuation that is not accounted for by time-invariant stock characteristics, by temporal patterns, or by recent price movements.

Table \ref{tab:pre_market_content} repeats my analysis using measures of emotions disaggregated between messages that convey earnings or trade-related information (``finance") and those that provide other types of information (``chat") in Column (1) and Column (2). I observe that emotions derived from messages specifically mentioning firm fundamentals and earnings generally present smaller point estimates for emotions with negative valence. This indicates that seemingly irrelevant messages carry important information that I capture by extracting emotions. For example, a user might post ":) :)" in response to news events or other posts, which clearly express their excitement but do not include any direct information about the stock beyond their emotional state. Furthermore, when comparing Column (3) to Column (4), I note that both messages containing original information and those disseminating existing information contribute to my findings. Notably, in the context of messages that disseminate existing information, only the levels of fear and happiness show statistically significant deviations from the baseline (neutral) level.

\subsection{Emotions and Persistence}

Table \ref{tab:pre_market_reversal} explores the impact on emotions before the market opens today on returns the following four days. As expected, I find that the predictive power of investor emotions diminishes over time. Interestingly, I do find persistent negative impacts for sadness, suggesting its uniquely enduring influence on market dynamics. Furthermore, in Panel B, when I consolidate these emotional measures into a singular metric, valence\footnote{I follow \cite{breaban2018emotional} and define valence as:
\begin{equation}
    \text{Valence} = \text{Happy} - \text{Sad} - \text{Anger} - \text{Disgust} - \text{Fear}
\end{equation}}, I observe a reduction in both the predictive power (evidenced by lower $R^2$ values) and the statistical significance of the emotions. This outcome highlights the inherent value in examining granular emotions rather than relying on a one-dimensional measure like valence.

\begin{table}[htbp]\centering
\def\sym#1{\ifmmode^{#1}\else\(^{#1}\)\fi}
\begin{threeparttable}
\caption{Empirical tests of Lab Experiments on Emotions and Asset Pricing}\label{tab:tests}
\footnotesize
\begin{tabular}[l]{|p{2in}|p{1in}|p{0.6in}|p{0.7in}|p{1in}|p{0.6in}|p{0.1in}|}\hline 
  & & & & & &\\ 
 Finding & Source & Dependent Variable & Sample & Test  & Result & \\ \hline 
  & & & & & &\\ 
 \textbf{Panel A: Directly Testable} & & & && & \\ \hline 
 & & & &&  \\[\dimexpr-\normalbaselineskip+2pt]
 & & & & & \\[\dimexpr-\normalbaselineskip+2pt]
\emph{Finding I:}  Average trader fear before the market  opens is negatively correlated with subsequent price movements.& \cite{breaban2018emotional} &  Return & Pre-market & $\rho_{y,fear}<0$   & $-0.0045$\sym{***}  & $\checkmark$ \\ \hline 
 & & & & & \\[\dimexpr-\normalbaselineskip+2pt]
\emph{Finding II:}  Fear and anger are negatively, while happiness is positively correlated with price movements.& \cite{hargreaves2011emotions}, \cite{breaban2018emotional} &  Return & Market & $\rho_{y,fear}<0$,  $\rho_{y,anger} < 0$, $\rho_{y,happy}>0$   & $-0.1879$\sym{***}, $-0.1684$\sym{***}, $0.2337$\sym{***}  & $\checkmark$ $\checkmark$ $\checkmark$ \\ \hline
 & & & & & &\\ 
 \textbf{Panel B: Indirectly Testable} & & & & & &\\ \hline 
 & & & & & & \\[\dimexpr-\normalbaselineskip+2pt]
 \emph{Finding III:}  Traders with higher levels of fear (valence) make more (less) loss-averse decisions. \emph{Alternative III:} Average trader fear (valence) before the market opens is positively (negatively) correlated with the ease of valuation. & \cite{breaban2018emotional} &  Market Cap & Pre-market & $\rho_{y,fear} > 0$, $\rho_{y,valence} < 0$ & $ 0.0151$\sym{***}, $--0.1717$\sym{***}  & $\checkmark$ $\checkmark$ \\ \hline 
  & & & & & & \\[\dimexpr-\normalbaselineskip+2pt]
 & & & & & & \\[\dimexpr-\normalbaselineskip+2pt]
\emph{Finding IV:}  Cash holdings are negatively correlated with fear. \emph{Alternative IV: Trading volume is positively correlated with fear.} & \cite{breaban2018emotional}& Volume & Market & $\rho_{y,fear} > 0$ & $0.0161$\sym{***}  & $\checkmark$ \\ \hline \hline 
\end{tabular}
\begin{tablenotes}
\footnotesize 
\item Notes: This table investigates whether the findings of controlled laboratory experiments that relate investor emotions to trading behavior replicate my observational data. \sym{*} \(p<0.10\), \sym{**} \(p<0.05\), \sym{***} \(p<0.01\). Continuous variables winsorized at the 0.1\% and 99.9\% level. Aside from Finding III, variables are demeaned by firm and date.
\end{tablenotes}
\end{threeparttable}
\end{table}

\begin{table}[htbp]\centering
\small
\begin{threeparttable}
\def\sym#1{\ifmmode^{#1}\else\(^{#1}\)\fi}
\caption{Pre-Market Emotions and Price Movements \label{tab:pre_market}}
\begin{tabular}[lcccc]{p{1.5in}p{1in}p{1in}p{1in}p{1in}}

\hline\hline
 &                     &                     &                                    \\[\dimexpr-\normalbaselineskip+2pt]
                    &\multicolumn{1}{c}{(1)}&\multicolumn{1}{c}{(2)}&\multicolumn{1}{c}{(3)}&\multicolumn{1}{c}{(4)}\\
\hline
                     &                     &                     &                    &                       \\[\dimexpr-\normalbaselineskip+2pt]
Happy               &                     &      0.0052\sym{***}&      0.0007         &      0.0430\sym{***}\\
                    &                     &    (0.0007)         &    (0.0008)         &    (0.0064)         \\
                     &                     &                     &                    &                       \\[\dimexpr-\normalbaselineskip+2pt]
Sad                 &                     &     -0.0242\sym{***}&     -0.0096\sym{***}&     -0.0341         \\
                    &                     &    (0.0031)         &    (0.0037)         &    (0.0279)         \\
                     &                     &                     &                    &                       \\[\dimexpr-\normalbaselineskip+2pt]
Fear                &                     &     -0.0064\sym{***}&     -0.0047\sym{**} &      0.0094         \\
                    &                     &    (0.0018)         &    (0.0019)         &    (0.0156)         \\
                     &                     &                     &                    &                       \\[\dimexpr-\normalbaselineskip+2pt]
Disgust             &                     &     -0.0041         &     -0.0042         &     -0.0233         \\
                    &                     &    (0.0033)         &    (0.0042)         &    (0.0239)         \\
                     &                     &                     &                    &                       \\[\dimexpr-\normalbaselineskip+2pt]
Anger               &                     &     -0.0050         &     -0.0036         &     -0.0062         \\
                    &                     &    (0.0055)         &    (0.0074)         &    (0.0423)         \\
                     &                     &                     &                    &                       \\[\dimexpr-\normalbaselineskip+2pt]
Surprise            &                     &     -0.0068\sym{***}&     -0.0057\sym{***}&      0.0267\sym{*}  \\
                    &                     &    (0.0017)         &    (0.0021)         &    (0.0148)         \\
\hline
                     &                     &                     &                    &                       \\[\dimexpr-\normalbaselineskip+2pt]
S\&P Super Composite Firms      &                     &                     &           X         &                     \\
At least 100 messages&                     &                     &                     &           X         \\ \hline 
                     &                     &                     &                    &                       \\[\dimexpr-\normalbaselineskip+2pt]
$\sigma_{y, within}$&      0.0475         &      0.0475         &      0.0308         &      0.0669         \\
Observations        & 454,328        & 454,328        & 163,011        &  53,803        \\
$R^2$               &      0.1047         &      0.1052         &      0.1443         &      0.1742         \\
\hline\hline
\end{tabular}
\begin{tablenotes}
\footnotesize 
\item Notes: This table considers the relationship between pre-market emotions and daily price movements. The dependent variable is daily returns, computed as the difference between the closing and the opening price, divided by the opening price. All specifications include firm and date fixed effects, close-open returns, lag open-close returns, past 20 trading days return, and volatility (as defined in Variable Definitions). Robust standard errors clustered at the industry and date levels are in parentheses. I use the Fama-French 12-industry classification. \sym{*} \(p<0.10\), \sym{**} \(p<0.05\), \sym{***} \(p<0.01\). Continuous variables winsorized at the 0.1\% and 99.9\% level to mitigate the impact of outliers. I report the within-firm, detrended (demeaned) standard deviation of the dependent variable. 
\end{tablenotes}
\end{threeparttable}
\end{table}

\begin{table}[htbp]\centering
\small
\begin{threeparttable}
\def\sym#1{\ifmmode^{#1}\else\(^{#1}\)\fi}
\caption{Pre-Market Emotions, Message Content and Price Movements \label{tab:pre_market_content}}
\begin{tabular}[lcccc]{p{1.5in}p{1in}p{1in}p{1in}p{1in}}
\hline\hline
 &                     &                     &                     &                \\[\dimexpr-\normalbaselineskip+2pt]
                    &\multicolumn{1}{c}{(1)}&\multicolumn{1}{c}{(2)}&\multicolumn{1}{c}{(3)}&\multicolumn{1}{c}{(4)}\\
                    & \multicolumn{2}{c}{\uline{Chat Type}} & \multicolumn{2}{c}{\uline{Information Type}}  \\ 
                    &\multicolumn{1}{c}{Chat}&\multicolumn{1}{c}{Finance}&\multicolumn{1}{c}{Disseminating}&\multicolumn{1}{c}{Original}\\
\hline
 &                     &                     &                     &                \\[\dimexpr-\normalbaselineskip+2pt]
Happy & 0.0028\sym{***} & 0.0025\sym{***} & 0.0030\sym{***} & 0.0027\sym{***} \\
 & (0.0004) & (0.0006) & (0.0005) & (0.0005) \\
 &                     &                     &                     &                \\[\dimexpr-\normalbaselineskip+2pt]
Sad & -0.0063\sym{***} & -0.0155\sym{***} & -0.0045\sym{**}& -0.0104\sym{***} \\
 & (0.0016) & (0.0025) & (0.0020) & (0.0018) \\
 &                     &                     &                     &                \\[\dimexpr-\normalbaselineskip+2pt]
Fear & -0.0021\sym{*}& -0.0039\sym{***} & -0.0016 & -0.0043\sym{***} \\
 & (0.0012) & (0.0014) & (0.0012) & (0.0011) \\
 &                     &                     &                     &                \\[\dimexpr-\normalbaselineskip+2pt]
Disgust & -0.0037\sym{**}& -0.0020 & -0.0023 & -0.0047\sym{**}\\
 & (0.0016) & (0.0028) & (0.0022) & (0.0019) \\
 &                     &                     &                     &                \\[\dimexpr-\normalbaselineskip+2pt]
Anger & -0.0010 & -0.0061 & -0.0044 & -0.0041 \\
 & (0.0026) & (0.0044) & (0.0035) & (0.0029) \\
 &                     &                     &                     &                \\[\dimexpr-\normalbaselineskip+2pt]
Surprise & -0.0031\sym{***} & -0.0027\sym{**}& 0.0005 & -0.0037\sym{***} \\
 & (0.0008) & (0.0013) & (0.0010) & (0.0009) \\
\hline
 &                     &                     &                     &                \\[\dimexpr-\normalbaselineskip+2pt]
$\sigma_{y, within}$&      0.0484         &      0.0475         &      0.0475         &      0.0487         \\
Observations        & 419,232        & 454,153        & 448,054        & 413,359        \\
$R^2$               &      0.1078         &      0.1050         &      0.1052         &      0.1098         \\
\hline\hline
\end{tabular}
\begin{tablenotes}
\footnotesize 
\item Notes: This table considers the relationship between pre-market emotions disaggregated by information content and information type and daily price movements. The dependent variable is daily returns, computed as the difference between the closing and the opening price, divided by the opening price. All specifications include firm and date fixed effects, close-open returns, lag open-close returns, past 20 trading days return, and volatility (as defined in Variable Definitions). Robust standard errors clustered at the industry and date levels are in parentheses. I use the Fama-French 12-industry classification. \sym{*} \(p<0.10\), \sym{**} \(p<0.05\), \sym{\sym{***}} \(p<0.01\). Continuous variables winsorized at the 0.1\% and 99.9\% level to mitigate the impact of outliers. I report the within-firm, detrended (demeaned) standard deviation of the dependent variable. 
\end{tablenotes}
\end{threeparttable}
\end{table}

\begin{table}[htbp]\centering
\small
\begin{threeparttable}
\def\sym#1{\ifmmode^{#1}\else\(^{#1}\)\fi}
\caption{Pre-Market Emotions and Subsequent Price Movements \label{tab:pre_market_reversal}}
\begin{tabular}[lcccc]{p{1.5in}p{1in}p{1in}p{1in}p{1in}}
\hline\hline
                    &\multicolumn{1}{c}{(1)}&\multicolumn{1}{c}{(2)}&\multicolumn{1}{c}{(3)}&\multicolumn{1}{c}{(4)}\\
                    &\multicolumn{1}{c}{Ret$_{t+1}$}&\multicolumn{1}{c}{Ret$_{t+2}$}&\multicolumn{1}{c}{Ret$_{t+3}$}&\multicolumn{1}{c}{Ret$_{t+4}$}\\
\hline
 &                     &                     &                     &                \\[\dimexpr-\normalbaselineskip+2pt]
\multicolumn{4}{l}{\textbf{Panel A: Granular Emotions}} \\
 &                     &                     &                     &                \\[\dimexpr-\normalbaselineskip+2pt]
Happy               &      0.0004         &      0.0002         &     -0.0003         &      0.0005         \\
                    &    (0.0007)         &    (0.0007)         &    (0.0006)         &    (0.0006)         \\
 &                     &                     &                     &                \\[\dimexpr-\normalbaselineskip+2pt]
Sad                 &     -0.0090\sym{***}&     -0.0060\sym{**} &     -0.0050\sym{*}  &     -0.0021         \\
                    &    (0.0029)         &    (0.0028)         &    (0.0028)         &    (0.0028)         \\
 &                     &                     &                     &                \\[\dimexpr-\normalbaselineskip+2pt]
Fear                &      0.0026         &      0.0010         &      0.0039\sym{**} &     -0.0002         \\
                    &    (0.0017)         &    (0.0017)         &    (0.0017)         &    (0.0016)         \\
 &                     &                     &                     &                \\[\dimexpr-\normalbaselineskip+2pt]
Disgust             &      0.0069\sym{**} &      0.0037         &     -0.0036         &      0.0062\sym{*}  \\
                    &    (0.0033)         &    (0.0032)         &    (0.0032)         &    (0.0032)         \\
 &                     &                     &                     &                \\[\dimexpr-\normalbaselineskip+2pt]
Anger               &      0.0025         &      0.0034         &      0.0096\sym{*}  &      0.0024         \\
                    &    (0.0055)         &    (0.0055)         &    (0.0055)         &    (0.0053)         \\
 &                     &                     &                     &                \\[\dimexpr-\normalbaselineskip+2pt]
Surprise            &     -0.0040\sym{**} &     -0.0010         &     -0.0025         &     -0.0010         \\
                    &    (0.0016)         &    (0.0015)         &    (0.0015)         &    (0.0015)         \\ \hline 
 &                     &                     &                     &                \\[\dimexpr-\normalbaselineskip+2pt]
$R^2$               &      0.1113         &      0.1141         &      0.1180         &      0.1160         \\ \hline 
 &                     &                     &                     &                \\[\dimexpr-\normalbaselineskip+2pt]
\multicolumn{4}{l}{\textbf{Panel B: Combined Emotions}} \\
 &                     &                     &                     &                \\[\dimexpr-\normalbaselineskip+2pt]
Valence &     -0.0002         &     -0.0001         &     -0.0004         &     -0.0002         \\
                    &    (0.0005)         &    (0.0005)         &    (0.0005)         &    (0.0005)         \\ \hline 
 &                     &                     &                     &                \\[\dimexpr-\normalbaselineskip+2pt]
$R^2$               &      0.1107         &      0.1140         &      0.1180         &      0.1160         \\                  
\hline
 &                     &                     &                     &                \\[\dimexpr-\normalbaselineskip+2pt]
$\sigma_{y, within}$&      0.0431         &      0.0418         &      0.0410         &      0.0405         \\
Observations        & 454,169       & 454,067       & 453,986       & 453,878       \\

\hline\hline
\end{tabular}
\begin{tablenotes}
\footnotesize 
\item Notes: This table investigates the predictive power of investor emotions on price movements one to four days into the future. The dependent variable is daily returns, computed as the difference between the closing and the opening price, divided by the opening price at $t+1$ through $t+4$. All specifications include firm and date fixed effects, close-open returns, open-close returns at $t-1$, past 20 trading days return and volatility (as defined in Variable Definitions). Robust standard errors clustered at the industry and date levels are in parentheses. I use the Fama-French 12-industry classification. \sym{*} \(p<0.10\), \sym{**} \(p<0.05\), \sym{***} \(p<0.01\). Continuous variables winsorized at the 0.1\% and 99.9\% level to mitigate the impact of outliers. I report the within-firm, detrended (demeaned) standard deviation of the dependent variable. 
\end{tablenotes}
\end{threeparttable}
\end{table}

\clearpage

\section{Additional Findings}\label{sec:additional}

The findings up to this point indicate that investor emotions, as extracted from social media, offer insightful information capable of forecasting daily stock price fluctuations. Nevertheless, the impact of these emotions is unlikely to be uniform. In this section, I delve into the heterogeneous effects linked to both stock and user characteristics, demonstrating that my results are not only robust but also intuitively reflect the underlying investor emotions.

\subsection{Heterogeneous Effects: Stock Characteristic}

The theoretical framework surrounding investor sentiment suggests that stocks that are younger, smaller, more volatile, unprofitable, do not pay dividends, and are in distress, exhibit a higher sensitivity to investor sentiment. On the other hand, stocks that resemble bonds in their characteristics tend to be less influenced by sentiment fluctuations, as discussed in \cite{baker2007investor}.

\subsubsection{Emotions, Volatility, and Liquidity}

To investigate if emotions mirror the behavior of sentiment, I interact my emotion variables with a dummy variable designed to encapsulate volatility and liquidity aspects, such as market capitalization and dollar trading volume. Consistent with theoretical insights on investor sentiment, I find larger point estimates for investor enthusiasm in firms characterized by smaller market capitalization (as seen in Column (4) of Table \ref{tab:pre_market_volat}).

\subsubsection{Emotions, Short Interest and Institutional Ownership}

I next interact my emotion variables with institutional ownership and short interest. I find larger impacts when short interest is higher (Column (5) of Table \ref{tab:pre_market_volat}). 

\subsection{Heterogeneous Effects: User Characteristic}

\cite{hong2004groups} demonstrate that a diverse group of intelligent decision-makers consistently outperforms a more homogenous group of highly skilled individuals. I explore this concept by categorizing the messages I have collected based on the traders' investment horizons (i.e., long-term vs. short-term), trading methodologies (i.e., value vs. technical), and levels of trading experience (i.e., amateur vs. professional).

In Table \ref{tab:pre_market_types}, I report the heterogeneity across user types and make several notable observations. First, consistent with the hypothesis on the benefits of diversity, I discover that the emotions expressed by homogeneous groups offer less predictive power regarding daily price fluctuations. Second, I observe that the correlation between investor emotions and daily stock movements remains consistent and statistically significant across the majority of my specifications.

\subsection{Sensitivity Analysis}\label{sec:sensitivity}
Next, I show that my results are robust and capture investor emotions intuitively.

\subsubsection{Alternative Classification}

I use alternative emotion variables using a model trained on Twitter emotion metadata collected by other researchers. I examine this Twitter-based model in Columns (1-3) of Table \ref{tab:pre_market_robustness}. My results continue to hold significance, lending robust support to the link between investor emotions and daily stock movements. \cite{vamossy2023emtract} provide a detailed explanation for the preference for the StockTwits model over the Twitter model.

\subsubsection{Alternative Weighting}

For my previous findings, I applied a weighting to messages based on log(1+followers). As a measure to test the validity of this approach, I explore a different weighting scheme. Specifically, I consider a scenario where I discard the initial weighting approach, thereby treating all messages with equal importance (as shown in Columns (4-6) of Table \ref{tab:pre_market_robustness}). My results remain mostly consistent even with these alternative specifications.

\subsubsection{Contemporaneous Emotions \& Prices}

To further validate that my emotion metrics accurately reflect investor emotions, I also run my primary regression using contemporaneous (within trading hours) emotion variables. Incorporating emotions measured during market hours into the analysis considerably enhances the model's robustness. The results presented in Table \ref{tab:market_ret} reinforce the accuracy of the happiness metric in capturing investor enthusiasm. Specifically, by including contemporaneous emotion variables, the $R^2$ of the primary regression experienced a significant increase of 8.46 percentage points, which constitutes a 75\% improvement in the model's explanatory power. This substantial improvement in $R^2$ underscores the added value of considering investor emotions within trading hours, affirming the strong link between positive market returns and elevated levels of enthusiasm among investors. Furthermore, the point estimates for these emotion variables increased by more than an order of magnitude, which supports the argument that these emotion metrics serve as reliable, multi-dimensional, and granular indicators of investor sentiment, especially during active trading periods. Nonetheless, I refrain from delving deeper into the contemporaneous effects, as they might be reactive rather than predictive.

\begin{sidewaystable}[htbp]\centering
\footnotesize
\begin{threeparttable}
\def\sym#1{\ifmmode^{#1}\else\(^{#1}\)\fi}
\caption{Pre-Market Emotions, Stock Characteristics and Price Movements \label{tab:pre_market_volat}}
\begin{tabular}[lcccc]{p{1in}p{0.7in}p{0.7in}p{0.7in}p{0.7in}p{0.7in}p{1in}}
\hline\hline
                    &\multicolumn{1}{c}{(1)}&\multicolumn{1}{c}{(2)}&\multicolumn{1}{c}{(3)}&\multicolumn{1}{c}{(4)}&\multicolumn{1}{c}{(5)}&\multicolumn{1}{c}{(6)}\\
                    & \multicolumn{6}{c}{\uline{Interaction Variable}} \\
                    &\multicolumn{1}{c}{None}&\multicolumn{1}{c}{\$ Volume$_{p25}$}&\multicolumn{1}{c}{Volatility$_{p75}$}&\multicolumn{1}{c}{Market Cap$_{p25}$}&\multicolumn{1}{c}{Short Interest$_{p75}$}&\multicolumn{1}{c}{Institutional$_{p25}$}\\
\hline
 &                     &                     &                     &       & &         \\[\dimexpr-\normalbaselineskip+2pt]
Happy               &      0.0048\sym{***}&      0.0057\sym{***}&      0.0043\sym{***}&      0.0059\sym{***}&      0.0040\sym{***}&      0.0040\sym{***}\\
                    &    (0.0007)         &    (0.0007)         &    (0.0007)         &    (0.0007)         &    (0.0008)         &    (0.0011)         \\
 &                     &                     &                     &       & &         \\[\dimexpr-\normalbaselineskip+2pt]
Sad                 &     -0.0238\sym{***}&     -0.0175\sym{***}&     -0.0232\sym{***}&     -0.0186\sym{***}&     -0.0224\sym{***}&     -0.0267\sym{***}\\
                    &    (0.0031)         &    (0.0032)         &    (0.0031)         &    (0.0032)         &    (0.0036)         &    (0.0044)         \\
 &                     &                     &                     &       & &         \\[\dimexpr-\normalbaselineskip+2pt]
Anger               &     -0.0068         &      0.0044         &     -0.0048         &     -0.0061         &     -0.0080         &     -0.0129         \\
                    &    (0.0055)         &    (0.0059)         &    (0.0054)         &    (0.0059)         &    (0.0064)         &    (0.0081)         \\
 &                     &                     &                     &       & &         \\[\dimexpr-\normalbaselineskip+2pt]
Surprise            &     -0.0070\sym{***}&     -0.0065\sym{***}&     -0.0052\sym{***}&     -0.0060\sym{***}&     -0.0059\sym{***}&     -0.0061\sym{**} \\
                    &    (0.0017)         &    (0.0018)         &    (0.0017)         &    (0.0017)         &    (0.0020)         &    (0.0025)         \\
 &                     &                     &                     &       & &         \\[\dimexpr-\normalbaselineskip+2pt]
Disgust             &     -0.0029         &     -0.0110\sym{***}&     -0.0039         &     -0.0068\sym{**} &      0.0017         &     -0.0036         \\
                    &    (0.0033)         &    (0.0034)         &    (0.0032)         &    (0.0034)         &    (0.0040)         &    (0.0048)         \\
 &                     &                     &                     &       & &         \\[\dimexpr-\normalbaselineskip+2pt]
Fear                &     -0.0070\sym{***}&     -0.0050\sym{***}&     -0.0065\sym{***}&     -0.0060\sym{***}&     -0.0064\sym{***}&     -0.0090\sym{***}\\
                    &    (0.0018)         &    (0.0018)         &    (0.0018)         &    (0.0018)         &    (0.0021)         &    (0.0025)         \\
 &                     &                     &                     &       & &         \\[\dimexpr-\normalbaselineskip+2pt]
Happy $\times$ IV             &                     &     -0.0020         &      0.0022         &     -0.0032\sym{*}  &      0.0027\sym{*}  &     -0.0006         \\
                    &                     &    (0.0016)         &    (0.0018)         &    (0.0017)         &    (0.0014)         &    (0.0027)         \\
 &                     &                     &                     &       & &         \\[\dimexpr-\normalbaselineskip+2pt]
Sad $\times$ IV               &                     &     -0.0250\sym{***}&     -0.0044         &     -0.0190\sym{**} &     -0.0054         &     -0.0171         \\
                    &                     &    (0.0081)         &    (0.0088)         &    (0.0079)         &    (0.0065)         &    (0.0110)         \\
 &                     &                     &                     &       & &         \\[\dimexpr-\normalbaselineskip+2pt]
Anger $\times$ IV             &                     &     -0.0294\sym{**} &     -0.0015         &     -0.0013         &      0.0018         &      0.0165         \\
                    &                     &    (0.0132)         &    (0.0144)         &    (0.0124)         &    (0.0114)         &    (0.0202)         \\
 &                     &                     &                     &       & &         \\[\dimexpr-\normalbaselineskip+2pt]
Surprise $\times$ IV          &                     &     -0.0011         &     -0.0064         &     -0.0031         &     -0.0048         &      0.0038         \\
                    &                     &    (0.0042)         &    (0.0046)         &    (0.0040)         &    (0.0034)         &    (0.0061)         \\
 &                     &                     &                     &       & &         \\[\dimexpr-\normalbaselineskip+2pt]
Disgust $\times$ IV           &                     &      0.0217\sym{***}&     -0.0008         &      0.0090         &     -0.0145\sym{**} &      0.0009         \\
                    &                     &    (0.0080)         &    (0.0089)         &    (0.0076)         &    (0.0069)         &    (0.0114)         \\
 &                     &                     &                     &       & &         \\[\dimexpr-\normalbaselineskip+2pt]
Fear $\times$ IV              &                     &     -0.0065         &     -0.0011         &     -0.0037         &     -0.0033         &      0.0118\sym{*}  \\
                    &                     &    (0.0049)         &    (0.0054)         &    (0.0048)         &    (0.0038)         &    (0.0071)         \\
 &                     &                     &                     &       & &         \\[\dimexpr-\normalbaselineskip+2pt]
IV                  &                     &      0.0046\sym{***}&      0.0001         &      0.0072\sym{***}&      0.0020\sym{***}&     -0.0011         \\
                    &                     &    (0.0007)         &    (0.0008)         &    (0.0008)         &    (0.0006)         &    (0.0013)         \\

\hline
 &                     &                     &                     &       & &         \\[\dimexpr-\normalbaselineskip+2pt]
$\sigma_{y, within}$&      0.0477         &      0.0475         &      0.0475         &      0.0477         &      0.0476         &      0.0451         \\
Observations        & 467,506        & 454,364        & 454,328        & 467,506        & 467,196        & 189,588        \\
$R^2$               &      0.1053         &      0.1053         &      0.1052         &      0.1061         &      0.1055         &      0.1096         \\
\hline\hline
\end{tabular}
\begin{tablenotes}
\footnotesize 
\item Notes: This table considers the relationship between pre-market emotions, stock characteristics, and daily price movements. The dependent variable is daily returns, computed as the difference between the closing and the opening price, divided by the opening price. The interaction variables (IV) are dummy variables for \$ trading Volume (Column (2)), volatility (Column (3)), market cap (Column (4)), short interest (Column (5)), and institutional ownership (Column (6)). All specifications include firm and date fixed effects, close-open returns, lag open-close returns, and past 20 trading days returns. Robust standard errors clustered at the industry and date levels are in parentheses. I use the Fama-French 12-industry classification. \sym{*} \(p<0.10\), \sym{**} \(p<0.05\), \sym{***} \(p<0.01\). Continuous variables winsorized at the 0.1\% and 99.9\% level to mitigate the impact of outliers. I report the within-firm, detrended (demeaned) standard deviation of the dependent variable. 
\end{tablenotes}
\end{threeparttable}
\end{sidewaystable}

\begin{table}[htbp]\centering
\small
\begin{threeparttable}
\def\sym#1{\ifmmode^{#1}\else\(^{#1}\)\fi}
\caption{Pre-Market Emotions, Investor Types and Price Movements\label{tab:pre_market_types}}
\begin{tabular}{l*{6}{c}}
\hline\hline
                    &\multicolumn{1}{c}{(1)}&\multicolumn{1}{c}{(2)}&\multicolumn{1}{c}{(3)}&\multicolumn{1}{c}{(4)}&\multicolumn{1}{c}{(5)}&\multicolumn{1}{c}{(6)}\\
                  & \multicolumn{2}{c}{\underline{Trading Experience}} & \multicolumn{2}{c}{\underline{Trading Approach}} & \multicolumn{2}{c}{\underline{Investment Horizon}}  \\
                 &  Amateur & Professional & Fundamental & Technical & Short-Term & Long-Term \\ \hline
   &        &       &        &        &       &         \\[\dimexpr-\normalbaselineskip+2pt]
Happy               &      0.0010\sym{**} &      0.0009\sym{**} &      0.0018\sym{***}&      0.0012\sym{***}&      0.0018\sym{***}&      0.0016\sym{***}\\
                    &    (0.0004)         &    (0.0004)         &    (0.0004)         &    (0.0004)         &    (0.0004)         &    (0.0004)         \\
   &        &       &        &        &       &         \\[\dimexpr-\normalbaselineskip+2pt]
Sad                 &     -0.0064\sym{***}&     -0.0041\sym{**} &     -0.0029\sym{**} &     -0.0068\sym{***}&     -0.0058\sym{***}&     -0.0052\sym{***}\\
                    &    (0.0014)         &    (0.0016)         &    (0.0014)         &    (0.0014)         &    (0.0013)         &    (0.0015)         \\
   &        &       &        &        &       &         \\[\dimexpr-\normalbaselineskip+2pt]
Anger               &     -0.0030         &     -0.0051         &     -0.0039         &     -0.0065\sym{**} &     -0.0042\sym{*}  &     -0.0009         \\
                    &    (0.0024)         &    (0.0031)         &    (0.0025)         &    (0.0026)         &    (0.0025)         &    (0.0026)         \\
   &        &       &        &        &       &         \\[\dimexpr-\normalbaselineskip+2pt]
Surprise            &     -0.0026\sym{***}&     -0.0018\sym{**} &     -0.0012         &     -0.0019\sym{**} &     -0.0011         &     -0.0009         \\
                    &    (0.0008)         &    (0.0009)         &    (0.0008)         &    (0.0009)         &    (0.0008)         &    (0.0008)         \\
   &        &       &        &        &       &         \\[\dimexpr-\normalbaselineskip+2pt]
Disgust             &     -0.0042\sym{***}&     -0.0006         &     -0.0001         &     -0.0027\sym{*}  &     -0.0032\sym{**} &     -0.0035\sym{**} \\
                    &    (0.0015)         &    (0.0020)         &    (0.0016)         &    (0.0016)         &    (0.0016)         &    (0.0017)         \\
   &        &       &        &        &       &         \\[\dimexpr-\normalbaselineskip+2pt]
Fear                &     -0.0021\sym{**} &     -0.0022\sym{**} &     -0.0028\sym{***}&     -0.0007         &     -0.0007         &     -0.0018\sym{*}  \\
                    &    (0.0009)         &    (0.0010)         &    (0.0009)         &    (0.0009)         &    (0.0009)         &    (0.0009)         \\
\hline
Observations        & 369,668      & 385,533      & 378,307      & 385,460      & 397,150      & 410,853      \\
$R^2$               &      0.1126         &      0.1051         &      0.1063         &      0.1058         &      0.1046         &      0.1071         \\
\hline\hline
\end{tabular}
\begin{tablenotes}
\footnotesize 
\item Notes: This table considers the relationship between pre-market emotions, user characteristics, and daily price movements. The dependent variable is daily returns, computed as the difference between the closing and the opening price, divided by the opening price. All specifications include firm and date fixed effects, close-open returns, lag open-close returns, past 20 trading days return, and volatility (as defined in Variable Definitions). Robust standard errors clustered at the industry and date levels are in parentheses. I use the Fama-French 12-industry classification. \sym{*} \(p<0.10\), \sym{**} \(p<0.05\), \sym{***} \(p<0.01\). Continuous variables winsorized at the 0.1\% and 99.9\% level to mitigate the impact of outliers. I report the within-firm, detrended (demeaned) standard deviation of the dependent variable. 
\end{tablenotes}
\end{threeparttable}
\end{table}

\begin{table}[htbp]\centering
\small
\begin{threeparttable}
\def\sym#1{\ifmmode^{#1}\else\(^{#1}\)\fi}
\caption{Robustness: Pre-Market Emotions and Price Movements \label{tab:pre_market_robustness}}
\begin{tabular}{l*{6}{c}}
\hline\hline
                    &\multicolumn{1}{c}{(1)}&\multicolumn{1}{c}{(2)}&\multicolumn{1}{c}{(3)}&\multicolumn{1}{c}{(4)}&\multicolumn{1}{c}{(5)}&\multicolumn{1}{c}{(6)}\\
                   &\multicolumn{3}{c}{Twitter Model}&\multicolumn{3}{c}{Alternative Weighting}\\
\hline
 &                     &                     &                     &   & &             \\[\dimexpr-\normalbaselineskip+2pt]
Happy               &      0.0025\sym{***}&      0.0011         &      0.0086         &      0.0055\sym{***}&      0.0006         &      0.0497\sym{***}\\
                    &    (0.0009)         &    (0.0010)         &    (0.0103)         &    (0.0007)         &    (0.0008)         &    (0.0075)         \\
 &                     &                     &                     &   & &             \\[\dimexpr-\normalbaselineskip+2pt]
Sad                 &     -0.0207\sym{***}&     -0.0084\sym{***}&     -0.0602\sym{***}&     -0.0266\sym{***}&     -0.0077\sym{**} &     -0.0361         \\
                    &    (0.0017)         &    (0.0017)         &    (0.0215)         &    (0.0031)         &    (0.0036)         &    (0.0286)         \\
 &                     &                     &                     &   & &             \\[\dimexpr-\normalbaselineskip+2pt]
Anger               &     -0.0116\sym{***}&     -0.0063\sym{**} &     -0.0265         &     -0.0088         &     -0.0069         &     -0.0243         \\
                    &    (0.0023)         &    (0.0029)         &    (0.0179)         &    (0.0055)         &    (0.0071)         &    (0.0430)         \\
 &                     &                     &                     &   & &             \\[\dimexpr-\normalbaselineskip+2pt]
Surprise            &      0.0011         &      0.0009         &     -0.0017         &     -0.0086\sym{***}&     -0.0057\sym{***}&      0.0411\sym{***}\\
                    &    (0.0015)         &    (0.0016)         &    (0.0202)         &    (0.0017)         &    (0.0020)         &    (0.0157)         \\
 &                     &                     &                     &   & &             \\[\dimexpr-\normalbaselineskip+2pt]
Disgust             &      0.0131         &      0.0003         &     -0.1245         &     -0.0028         &     -0.0033         &     -0.0081         \\
                    &    (0.0158)         &    (0.0156)         &    (0.1272)         &    (0.0033)         &    (0.0040)         &    (0.0243)         \\
 &                     &                     &                     &   & &             \\[\dimexpr-\normalbaselineskip+2pt]
Fear                &     -0.0040\sym{**} &     -0.0055\sym{***}&      0.0267         &     -0.0068\sym{***}&     -0.0049\sym{**} &      0.0183         \\
                    &    (0.0016)         &    (0.0017)         &    (0.0163)         &    (0.0019)         &    (0.0020)         &    (0.0168)         \\
\hline
 &                     &                     &                     &   & &             \\[\dimexpr-\normalbaselineskip+2pt]
S\&P Super Composite Firms      &                     &           X         &                     &                     &           X         &                     \\
At least 100 messages&                     &                     &           X         &                     &                     &           X         \\ \hline 
 &                     &                     &                     &   & &             \\[\dimexpr-\normalbaselineskip+2pt]
$\sigma_{y, within}$&      0.0481         &      0.0309         &      0.0673         &      0.0481         &      0.0309         &      0.0673         \\
Observations        & 454,328        & 163,011        &  53,803        & 454,328        & 163,011        &  53,803        \\
$R^2$               &      0.1052         &      0.1444         &      0.1735         &      0.1053         &      0.1443         &      0.1743         \\
\hline\hline
\end{tabular}
\begin{tablenotes}
\footnotesize 
\item Notes: This table considers the robustness of the relationship between pre-market emotions and daily price movements. Columns (1-3) use emotion variables constructed based on a model trained on Twitter emotion data compiled by other researchers. Columns (4-6) abandon the weighting scheme used in my analysis and weigh each posts equally. The dependent variable is daily returns, computed as the difference between the closing and the opening price, divided by the opening price. All specifications include firm and date fixed effects, close-open returns, lag open-close returns, past 20 trading days return, and volatility (as defined in Variable Definitions). Robust standard errors clustered at the industry and date levels are in parentheses. I use the Fama-French 12-industry classification. \sym{*} \(p<0.10\), \sym{**} \(p<0.05\), \sym{***} \(p<0.01\). Continuous variables winsorized at the 0.1\% and 99.9\% level to mitigate the impact of outliers. I report the within-firm, detrended (demeaned) standard deviation of the dependent variable. 
\end{tablenotes}
\end{threeparttable}
\end{table}

\begin{table}[htbp]\centering
\small
\begin{threeparttable}
\def\sym#1{\ifmmode^{#1}\else\(^{#1}\)\fi}
\caption{Market Emotions and Price Movements \label{tab:market_ret}}
\begin{tabular}[lcccc]{p{1.5in}p{1in}p{1in}p{1in}p{1in}}
\hline\hline
 &                     &                     &                     &                \\[\dimexpr-\normalbaselineskip+2pt]
                    &\multicolumn{1}{c}{(1)}&\multicolumn{1}{c}{(2)}&\multicolumn{1}{c}{(3)}&\multicolumn{1}{c}{(4)}\\
\hline
 &                     &                     &                     &                                     \\[\dimexpr-\normalbaselineskip+2pt]
Happy               &                     &      0.0821\sym{***}&      0.0547\sym{***}&      0.2535\sym{***}\\
                    &                     &    (0.0008)         &    (0.0008)         &    (0.0064)         \\
 &                     &                     &                     &                                     \\[\dimexpr-\normalbaselineskip+2pt]
Sad                 &                     &     -0.1892\sym{***}&     -0.1102\sym{***}&     -0.8765\sym{***}\\
                    &                     &    (0.0027)         &    (0.0033)         &    (0.0220)         \\
 &                     &                     &                     &                                     \\[\dimexpr-\normalbaselineskip+2pt]
Fear                &                     &     -0.1003\sym{***}&     -0.0507\sym{***}&     -0.4925\sym{***}\\
                    &                     &    (0.0018)         &    (0.0020)         &    (0.0145)         \\
 &                     &                     &                     &                                     \\[\dimexpr-\normalbaselineskip+2pt]
Surprise            &                     &     -0.0619\sym{***}&     -0.0456\sym{***}&     -0.0777\sym{***}\\
                    &                     &    (0.0017)         &    (0.0021)         &    (0.0141)         \\
 &                     &                     &                     &                                     \\[\dimexpr-\normalbaselineskip+2pt]
Disgust             &                     &     -0.1173\sym{***}&     -0.0500\sym{***}&     -0.3808\sym{***}\\
                    &                     &    (0.0029)         &    (0.0036)         &    (0.0214)         \\
 &                     &                     &                     &                                     \\[\dimexpr-\normalbaselineskip+2pt]
Anger               &                     &     -0.0677\sym{***}&     -0.0354\sym{***}&     -0.0539\sym{*}  \\
                    &                     &    (0.0044)         &    (0.0058)         &    (0.0321)         \\
\hline
                     &                     &                     &                    &                       \\[\dimexpr-\normalbaselineskip+2pt]
S\&P Super Composite Firms&                     &                     &           X         &                     \\
At least 100 messages&                     &                     &                     &           X         \\ \hline 
                     &                     &                     &                    &                       \\[\dimexpr-\normalbaselineskip+2pt]
$\sigma_{y, within}$&      0.0539         &      0.0539         &      0.0350         &      0.0773         \\
Observations        & 476,802         & 476,801         & 160,354         &  78,268         \\
$R^2$               &      0.1129         &      0.1975         &      0.2292         &      0.3389         \\
\hline\hline
\end{tabular}
\begin{tablenotes}
\footnotesize 
\item Notes: This table considers the relationship between emotions posted during market hours and daily price movements. The dependent variable is daily returns, computed as the difference between the closing and the opening price, divided by the opening price. All specifications include firm and date fixed effects, close-open returns, lag open-close returns, past 20 trading days return, and volatility (as defined in Variable Definitions). Robust standard errors clustered at the industry and date levels are in parentheses. I use the Fama-French 12-industry classification. \sym{*} \(p<0.10\), \sym{**} \(p<0.05\), \sym{***} \(p<0.01\). Continuous variables winsorized at the 0.1\% and 99.9\% level to mitigate the impact of outliers. I report the within-firm, detrended (demeaned) standard deviation of the dependent variable. 
\end{tablenotes}
\end{threeparttable}
\end{table}

\clearpage 

\section{Conclusion} \label{sec:conclusion}

This work contributes to the literature on investor emotions and market dynamics, corroborating the results from controlled laboratory experiments. I have shown that investor emotions are significant predictors of daily stock price movements, with their influence being more pronounced when accompanied by greater short interest, and lower levels of liquidity. Furthermore, the persistence of sadness on subsequent returns and the insignificance of the combined valence metric underscores the importance of dissecting emotional states to gain a deeper and more accurate insight into how investor sentiments drive market movements.

\singlespacing
\nocite{*}
\bibliographystyle{plainnat}
\bibliography{bib}

\clearpage

\onehalfspacing

\appendix
\clearpage

\counterwithin{figure}{section}
\counterwithin{table}{section}
\addcontentsline{toc}{section}{Appendices}

\section*{Appendix}\label{app:app}

\section{Variable Definitions and Sources}

Table \ref{tab:variable_definitions} defines the variables used in the analyses. 

\begin{center}
\onehalfspacing
\footnotesize 
\begin{ThreePartTable}

\begin{longtable}[H]{p{1.5in}p{4in}p{0.75in}}
\caption{Variable Definitions}\label{tab:variable_definitions} \\

Variable & Definition & Source \\ \hline 
\endfirsthead

\multicolumn{3}{c}%
{{\bfseries \tablename\ \thetable{} -- continued from previous page}} \\
Variable & Definition & Source \\ \hline 
& & \\[\dimexpr-\normalbaselineskip+5pt]
\endhead

\endlastfoot
Open-Close Return & Difference between daily closing and open price, normalized by the open price. & CRSP \\
Close-Open Return & Difference between the previous day's closing price and current open price, normalized by the previous day's closing price. & CRSP \\
Market Cap$_{-1}$  & Natural logarithm of the market value of equity ($\log$(1+CSHOC$*$PRCCD)). & CRSP \\
Volatility & Standard deviation of daily returns from 183 days prior up to a day before. & CRSP \\
\$ Volume$_{-183,-1}$ & Arithmetic mean of the natural logarithm of market value of daily trading volume ($\log$(1+CSHTRD$*$(PRCCD + PRCOD)/2)) from 183 days prior up to a day before. & CRSP \\
Short Interest & Shares short divided by shares outstanding. Bi-weekly frequency. & Compustat \\
Institutional Ownership & Number of shares held by institutional investors scaled by total shares outstanding as of the quarter-end date & Thomson Reuters Institutional Holdings (13F) \\
Emotion & Each message is classified by a many-to-one deep learning model into one of the seven categories (i.e., neutral, happy, sad, anger, disgust, surprise, fear) so that the corresponding probabilities sum up to 1. For each emotion separately, I then take the weighted average of these probabilities from 4 pm the previous day until 9.29 am the trading day. The weights correspond to the number of followers of the user 1+$\log$(1 + \# of Followers).  & StockTwits \\
& & \\ 
 \hline \hline 
\end{longtable}
\end{ThreePartTable}
\end{center}

\section{StockTwits Activity \& Sample Distributions}

Table \ref{tab:twits_by_year} displays the descriptive statistics of StockTwits activity for my dataset, particularly focusing on the annual frequency distributions. There is a noticeable escalation in StockTwits usage throughout my study period, with the number of messages surging from 49,074 in 2010 to 10,584,254 by 2020. This trend underscores the burgeoning role of social media over the 2010s. Likewise, the scope of my analysis broadened significantly, capturing firm-day observations that increased from 1,660 in 2010 to 129,565 by 2020.

In addition, I examined how the distribution of messages and firm-days in my sample aligned with the Fama-French 12-industry classifications compared to the broader CRSP universe during the same timeframe. These findings, detailed in Table \ref{tab:fama_fench}, reveal that my dataset encompasses all 12 industries, with a pronounced emphasis on technology and healthcare sectors and a lesser representation of financial firms.

\begin{table}[htbp]\centering 
\small
\caption{Distribution of Posts by Calendar Year \label{tab:twits_by_year}}
\def\sym#1{\ifmmode^{#1}\else\(^{#1}\)\fi}
\begin{tabular}{l*{2}{c}}
\hline\hline
& & \\[\dimexpr-\normalbaselineskip+2pt]
                    &\multicolumn{1}{c}{(1)}&\multicolumn{1}{c}{(2)}\\
                    &\multicolumn{1}{c}{Firm-Day Observations} &\multicolumn{1}{c}{Posts} \\
\hline
& & \\[\dimexpr-\normalbaselineskip+2pt]
\hline
& & \\[\dimexpr-\normalbaselineskip+2pt]
2010                &        1,337         &       39,768         \\
                    &                     &                     \\
& & \\[\dimexpr-\normalbaselineskip+2pt]
2011                &        3,543         &      115,710         \\
                    &                     &                     \\
& & \\[\dimexpr-\normalbaselineskip+2pt]
2012                &        4,857         &      295,552         \\
                    &                     &                     \\
& & \\[\dimexpr-\normalbaselineskip+2pt]
2013                &        6,718         &      399,211         \\
                    &                     &                     \\
& & \\[\dimexpr-\normalbaselineskip+2pt]
2014                &       11,285         &      692,239         \\
                    &                     &                     \\
& & \\[\dimexpr-\normalbaselineskip+2pt]
2015                &       17,256         &      893,803         \\
                    &                     &                     \\
& & \\[\dimexpr-\normalbaselineskip+2pt]
2016                &       22,734         &     1,270,722         \\
                    &                     &                     \\
& & \\[\dimexpr-\normalbaselineskip+2pt]
2017                &       41,840         &     2,796,966         \\
                    &                     &                     \\
& & \\[\dimexpr-\normalbaselineskip+2pt]
2018                &       53,472         &     3,453,727         \\
                    &                     &                     \\
& & \\[\dimexpr-\normalbaselineskip+2pt]
2019                &       58,645         &     3,351,474         \\
                    &                     &                     \\
& & \\[\dimexpr-\normalbaselineskip+2pt]
2020                &      116,338         &     9,772,462         \\
                    &                     &                     \\
& & \\[\dimexpr-\normalbaselineskip+2pt]
2021                &      141,438         &    14,575,580         \\
                    &                     &                     \\ \hline 
& & \\[\dimexpr-\normalbaselineskip+2pt]
Total               &      479,463         &    37,657,214         \\
                    &                     &                     \\
\hline\hline
\end{tabular}
\end{table} 
\begin{sidewaystable}[htbp]
\centering
\begin{threeparttable}
\caption{Distribution of Posts Based on Fama-French 12-Industry Classification\label{tab:fama_fench}}
\small 
\def\sym#1{\ifmmode^{#1}\else\(^{#1}\)\fi}
\begin{tabular}{l*{3}{c}}
\hline\hline
& & & \\[\dimexpr-\normalbaselineskip+2pt]
                    &\multicolumn{1}{c}{(1)}&\multicolumn{1}{c}{(2)}&\multicolumn{1}{c}{(3)}\\
Fama-French industry code (12 industries)                    &         CRSP (\%)&         Posts (\%)& Firm-Day (\%) \\
\hline
& & & \\[\dimexpr-\normalbaselineskip+2pt]
Consumer NonDurables -- Food, Tobacco, Textiles, Apparel, Leather, Toys&        4.01&        2.73&        2.99\\
Consumer Durables -- Cars, TV's, Furniture, Household Appliances&        2.30&        7.60&        3.35\\
Manufacturing -- Machinery, Trucks, Planes, Off Furn, Paper, Com Printing&        8.58&        3.01&        5.04\\
Oil, Gas, and Coal Extraction and Products&        3.13&        1.56&        2.91\\
Chemicals and Allied Products&        2.38&        1.34&        1.61\\
Business Equipment -- Computers, Software, and Electronic Equipment&       15.01&       22.62&       21.82\\
Telephone and Television Transmission&        1.75&        1.11&        1.75\\
Utilities           &        2.50&        0.31&        0.86\\
Wholesale, Retail, and Some Services (Laundries, Repair Shops)&        7.48&        5.92&        7.58\\
Healthcare, Medical Equipment, and Drugs&       15.13&       31.96&       33.21\\
Finance             &       24.01&        4.21&        8.23\\
Other -- Mines, Constr, BldMt, Trans, Hotels, Bus Serv, Entertainment&       13.72&       17.63&       10.66\\
\hline
& & & \\[\dimexpr-\normalbaselineskip+2pt]
Observations        &    10,357,038&    36,518,413&      471,307\\
\hline\hline
\end{tabular}
\begin{tablenotes}
\footnotesize 
\item Notes: CRSP sample follows the same basic sample restrictions; active, traded on U.S. exchanges.
\end{tablenotes}
\end{threeparttable}
\end{sidewaystable}

\section{Examples of StockTwits Model Outputs}

\begin{table}[!h]\centering
\begin{threeparttable}
\small 
\caption{Examples of StockTwits Model Outputs}\label{tab:twit_examples}
\begin{tabular}{llllllll}
Text                                         & Neutral & Happy  & Sad    & Anger  & Disgust & Surprise & Fear   \\ \hline
&   &  &   &   &    &    &  \\[\dimexpr-\normalbaselineskip+2pt]
Financial markets have been uneventful.      & 79.2\%  & 1.9\%  & 5.3\%  & 1.7\%  & 7.4\%   & 1.3\%    & 3.1\%  \\
Today has been such a nice day :).           & 0.5\%   & 99.0\% & 0.1\%  & 0.0\%  & 0.0\%   & 0.2\%    & 0.1\%  \\
Been a long time since i felt so awful :(. & 0.4\%   & 4.1\%  & 81.0\% & 1.1\%  & 4.5\%   & 3.6\%    & 5.4\%  \\
You freaking idiots! Stop selling!!          & 0.6\%   & 1.2\%  & 3.5\%  & 63.8\% & 26.0\%  & 2.4\%    & 2.4\%  \\
These nasty politicians gotta go!            & 2.0\%   & 7.0\%  & 1.6\%  & 12.6\% & 63.8\%  & 0.9\%    & 12.0\% \\
WTf is going on rn??                         & 2.7\%   & 0.3\%  & 0.7\%  & 0.8\%  & 0.6\%   & 93.7\%   & 1.2\%  \\
Pretty choppy lately!                        & 22.5\%  & 11.9\% & 23.0\% & 3.9\%  & 6.4\%   & 1.9\%    & 30.4\% \\ \hline \hline 
\end{tabular}
\begin{tablenotes}
\footnotesize 
\item Notes: Examples of inputs (text) and outputs (i.e., neutral, happy, sad, anger, disgust, surprise, fear) for my StockTwits-based emotion classification model.
\end{tablenotes}
\end{threeparttable}
\end{table}

\end{document}